\documentclass[twocolumn]{aastex631}
\usepackage{lineno}
\submitjournal{ApJ}

\shorttitle{Mock Catalogs for High-z Surveys}
\shortauthors{J. Meng et al.}

\begin{document}

\title{Measuring galaxy abundance and clustering at high redshift
  from incomplete spectroscopic data: Tests on mock catalogs}

\correspondingauthor{Jiacheng Meng \& Cheng Li}
\email{meng.jiacheng@foxmail.com}
\email{cli2015@tsinghua.edu.cn}

\author[0000-0002-1974-905X]{Jiacheng Meng}
\affiliation{Department of Astronomy, Tsinghua University, Beijing 100084, China}

\author[0000-0002-8711-8970]{Cheng Li}
\affiliation{Department of Astronomy, Tsinghua University, Beijing 100084, China}

\author[0000-0001-5356-2419]{H.J. Mo}
\affiliation{Department of Astronomy, University of Massachusetts Amherst, MA 01003, USA}

\author[0000-0002-4597-5798]{Yangyao Chen}
\affiliation{Department of Astronomy, University of Science and Technology of China, Hefei Anhui 230026, China}

\author[0000-0002-3775-0484]{Kai Wang}
\affiliation{Kavli Institute for Astronomy and Astrophysics, Peking University, Beijing 100871, China}



\begin{abstract}

The number density and correlation function of galaxies 
  are two key quantities to characterize the distribution of the observed galaxy 
  population. High-$z$ spectroscopic surveys, which usually
  involve complex target selection and are incomplete in redshift sampling, 
  present both opportunities and challenges to measure
  these quantities reliably in the high-$z$ Universe. 
  Using realistic mock catalogs we show that target selection and 
  redshift incompleteness can lead to significantly biased
  results, especially due to the flux limit selection criteria. 
  We develop a new method to correct the flux limit effect, using information 
  provided by the parent photometric data from which the 
  spectroscopic sample is constructed. Our tests using realistic 
  mock samples show that the method is able to reproduce the 
  true stellar mass function and correlation function reliably. 
  Mock catalogs are constructed for the existing  
  zCOSMOS and VIPERS surveys, as well as for the 
  forthcoming PFS galaxy evolution survey. 
  The same set of mock samples are used to quantify the total variance 
  expected for different sample sizes. 
  We find that the total variance decreases very slowly when the 
  survey area reaches about 4 deg$^2$ for the abundance and about 8 deg$^2$ 
  for the clustering, indicating that the cosmic variance is no longer 
  the dominant source of error for PFS-like surveys. We also quantify 
  improvements expected in the PFS-like galaxy survey relative   
  to zCOSMOS and VIPERS surveys.
\end{abstract}



\section{Introduction} \label{sec:intro}

The two most basic functions that characterize the observed galaxy 
population in the universe are the luminosity/stellar mass function 
\citep[e.g.][]{Cole2001,Bell2003,Baldry2008,Li2009} 
and the spatial two-point correlation function 
\citep[e.g.][]{JingMoBoerner1998,Norberg2001,Zehavi2005,Li-06b,Abbas2006,Zehavi_etal2011}. 
The former measures the number density of galaxies as a function 
of their luminosity or stellar mass, while the latter describes
how strongly galaxies are clustered in space. 
In the current cold dark matter paradigm of structure formation 
\citep[][and references therein]{White1978,MoBoschWhite2010}, 
these two functions provide the key to understanding how galaxies 
form and evolve in the cosmic density field. Indeed, the 
observed luminosity/stellar mass function and correlation function have 
been widely used to constrain theoretical models 
\citep[e.g.][]{Cole2000,YangMoBosch2003,Zheng2005,LuYu_etal2011,Artale_etal2017,
Springel_etal2018,WechslerTinker2018}. 

The best way to obtain both the stellar mass function and 
correlation function is through the use of spectroscopic 
surveys of galaxies, where distances of individual galaxies can 
be estimated from spectroscopically-measured redshifts and 
stellar masses can be estimated from the spectra combined with 
multi-band photometry \citep[][]{Ilbert2006,Pozzetti2007,Boquien2019}. 
A lot of efforts have been made to measure the two functions 
using various redshift surveys. For example, in the low-$z$ 
universe, galaxy stellar mass function has been estimated from 
2-degree Field Galaxy Redshift Survey 
\citep[2dFGRS;][]{Cole2001}, Sloan Digital Sky Survey \citep[SDSS;][]{Li2009}
and Galaxy and Mass Assembly survey \citep[GAMA;][]{Baldry2012}, 
while the two 
point correlation function has been measured from
the Las Campanas Survey \citep[e.g.][]{JingMoBoerner1998}, 
2dFGRS \citep[e.g.][]{Madgwick2003}, 
SDSS \citep[e.g.][]{Li-06b,Zehavi_etal2011} and GAMA \citep[e.g.][]{Farrow2015}. 
At higher redshift, galaxy stellar 
mass function has been measured from DEEP2 Galaxy Redshift Survey \citep[DEEP2;][]{Bundy2003}, 
zCOSMOS \citep{Pozzetti2010}, VIMOS-VLT Deep Survey \citep[VVDS;][]{Pozzetti2007}, 
and The VIMOS Public Extragalactic Redshift Survey \citep[VIPERS;][]{Davidzon2013}, 
while the two-point correlation function has been estimated from DEEP2 
\citep{Coil2006}, zCOSMOS \citep[][]{Meneux2009,Torre2011}, 
VVDS \citep{Pollo2006} and VIPERS \citep{Mohammad2018,Marulli2013}. 

To facilitate comparisons with model predictions, statistical
measurements of the galaxy population, such as the stellar mass 
function and correlation function, are useful only when the 
samples used for the measurements are well defined. However, 
owing to observational limitations, real samples usually contain 
complex selection effects which make them incomplete in sampling 
the galaxy population we are interested in. Such incompleteness 
is particularly important for galaxies at high redshift 
\citep[][]{Ilbert2005,Pozzetti2010} where 
observational limitations are severer than in the low-$z$ universe. 
Generally there are three incompleteness effects that have to be 
carefully taken into account. The first one, referred to 
as the {\em sampling rate effect} in the following, is that only a fraction of 
galaxy targets selected from the parent photometric sample can be 
included in the final spectroscopic sample, . 
This effect is contributed by the incomplete sampling of galaxy targets 
selected for observations and the imperfect redshift determination 
of a target galaxy. For example, the overall redshift sampling rate is 
only about $55\%$ for the zCOSMOS-bright survey \citep{Knobel2012}
and about $50-70\%$ for the forthcoming Prime Focus Spectrograph galaxy 
evolution survey \citep[PFS,][]{Takada2014}, much lower than 
the sampling rate of nearly $100\%$ for the low-$z$ surveys such 
as SDSS \citep{York-00}.
The sampling rates in general are not uniform across the sky 
\citep[][]{Torre2011} 
and may depend on intrinsic properties of galaxies \citep[][]{Zucca2009}. 
The second is the {\em fiber collision effect}
(or {\em slit collision effect} in slit mask spectroscopy), 
which prevents a close pair of galaxies 
from being observed simultaneously, and so can bias 
measurements of clustering on small scales \citep[][]{Hawkins2003,Li-06a}. 
The last effect is caused by flux selection criteria of the 
galaxy sample, which may bias against low-mass red 
galaxies because of their low flux in the observing band \citep[][]{Meneux2008,Meneux2009,Marulli2013}. 
This effect is referred to as the {\em flux limit effect} in the following.
Observed samples are thus a biased sampling of the underlying 
galaxy population whose statistical properties are subjects 
of our interest. Clearly, effects of such bias need to be 
corrected. In practice, almost all high-$z$ spectroscopic surveys 
are based on deep photometric surveys with multi-waveband 
data \citep[][]{Laigle2016,Aihara2018} 
that can be used not only to obtain photometric redshift 
but also to estimate color, luminosity and stellar mass of 
individual galaxies \citep[][]{Muzzin2013,Laigle2016}. 
This information can be combined with the 
spectroscopic data to provide a more 
faithful representation of the targeted galaxy population.

Another limitation on current high redshift galaxy surveys 
is that the samples are relatively small so that cosmic variance
is a serious concern \citep[][]{Somerville2004,Driver2010,Moster2011}. 
To quantify cosmic variance and effects 
due to sample selection, the best way is to use realistic 
mock catalogs that follow the observational selection criteria closely. 

The goal of this paper is to develop methods  
that can be used to measure galaxy stellar mass function and 
correlation function from high-redshift surveys that are 
significantly incomplete in sampling. Our methods combine spectroscopic 
galaxies with those in the parent photometric survey 
to make full use of the information provided by the observational data. 
In particular, we emphasize that the flux limit effect leads to 
significant underestimation of the clustering of low-mass red 
galaxies at high redshift, an effect which has been largely ignored 
in previous studies. We develop a new method to correct for 
this effect. We calibrate and test our methods using detailed mock catalogs 
that mimic real observations with different selection criteria 
and completeness. These mock catalogs are also used to investigate 
errors in the galaxy stellar mass function and correlation 
function expected from samples of different sizes.
We apply our method to existing surveys, zCOSMOS 
and VIPERS, as well as the upcoming PFS galaxy evolution survey 
\citep[][]{Takada2014,PFS-galaxysurvey}. 
The same mock catalogs have been used in \citet{Wang-20}
to test a new group finding algorithm, developed specifically for identifying 
galaxy groups/clusters in high-$z$ galaxy surveys. 

The paper is organized as followed. In \S~\ref{sec:sim_and_model_mock} 
we describe the simulation, the galaxy formation model and 
methods used to construct mock catalogs. 
In \S~\ref{sec:flux_limit_effect}, we describe the 
{\em flux limit effect} caused by the selection criteria in high-$z$ 
survey on the measurements of correlation function, as well as methods to 
correct this effect. We then apply our mock catalogs and measuring 
methods to PFS-like surveys in \S~\ref{sec:mock_test}. 
We summarize our results in \S~\ref{sec:summary}.
Throughout the paper, we assume the WMAP5 cosmology 
\citep[][]{Dunkley2009,Komatsu2009} with 
the density parameter $\Omega_{\rm m}=0.258$ and the 
Hubble constant $h=0.72$.

\section{Simulation, galaxy model, surveys and mock catalogs} 
\label{sec:sim_and_model_mock}

In this section, we first introduce numerical simulation and 
the galaxy formation model we use to construct mock catalogs, 
and then describe our method for constructing mock
catalogs. Specifically, we will construct mock catalogs based on 
three surveys: the upcoming Subaru/PFS galaxy evolution survey
and two existing surveys --- zCOSMOS and VIPERS.
We emphasize, however, that our method is by no means limited to these surveys,
and that the basic methodology should be equally applicable to any
surveys with incompleteness produced by color selection and spectroscopic 
sampling. 

\subsection{The ELUCID simulation}
\label{sec:simulation} 

We use the ELUCID simulation carried out by \citet{Wang2016}
to construct the mock catalogs. The ELUCID is a large $N$-body numerical
simulation using $3072^3$ dark matter particles in 
a cubic box of 500 ${\rm Mpc/h}$ on one side. The mass of a dark matter 
particle is $3.088\times 10^{8}M_{\odot}/h$. The simulation assumes 
the WMAP5 cosmology, which is a flat $\Lambda$CDM universe with a matter 
density parameter of $\Omega_{\rm m}=0.258$ and a Hubble constant given by
$h\equiv H_{0}/(100\,{\rm km/s/Mpc})=0.72$. The simulation was run from 
redshift $z=100$ to $z=0$, and 100 snapshots were recorded between 
$z=18.4$ and $z=0$. Dark matter halos and their 
substructures containing more than 20 particles are identified with the 
friends-of-friends (FOF) and SUBFIND algorithms \citep{Springel2005}, respectively.
A more detailed description of the simulation can be found in 
\cite{Wang2016}. Here we only use halos in the simulation with masses larger 
than $10^{10}M_{\odot}/h$. However, this mass limit is insufficient to resolve star formation in 
low-mass halos. As an remedy we adopt the Monte Carlo method of \citet{Parkinson2008} to extend the 
merger trees to a halo mass limit, $10^{9}M_{\odot}/h$. All halos and sub-halos in the extended trees 
are populated with galaxies based on the galaxy formation model described in the next sub-section. 
More details of the merger-tree extension can be found in \citet{Yangyao2019}. 

\subsection{The galaxy model}
\label{sec:model}

\subsubsection{The empirical model of galaxies}
\begin{figure}
	\includegraphics[width=\columnwidth]{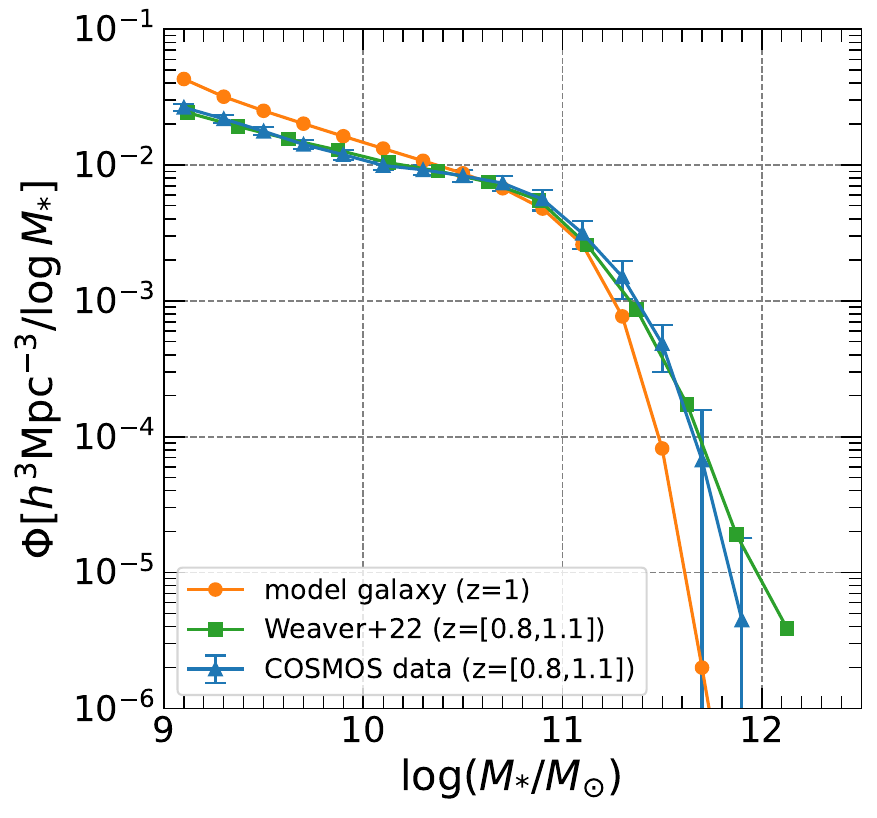}
    \caption{The galaxy stellar mass function at $z\sim 1$. The orange curve is 
    for model galaxies in the simulation and the blue curve is the 
    measurement obtained with the COSMOS2020 photometric sample, which we used 
    to calibrate the luminosity and color assignment for the model galaxies. 
    The green curve is the measurement from \citet{Weaver2022gsmf} based 
    on COSMOS2020 catalog.}
    \label{fig:model_mass}
\end{figure}

We populate dark matter (sub-)halos in the ELUCID simulation with
galaxies of different stellar masses and multi-band luminosities.
To this end, we first assume that each sub-halo hosts a galaxy 
and we assign a stellar mass to the galaxy using the galaxy formation model 
in \citet{Lu2014a, LuZ_etal2015}. The detail of this process is described in \cite{Yangyao2019}. 
In short, the 
model assumes that galaxies form at the center of dark matter 
halos, and the growth of central galaxies is a function of dark matter 
halo mass and redshift. A central galaxy becomes a satellite 
if its host halo is accreted into a more massive halo and becomes a sub-halo.
Thereafter the star formation rate of the galaxy (thus its growth) 
is suppressed due to some environmental quenching processes.   
The positions and velocities of galaxies are
determined by those of their host halos (for centrals) 
or sub-halos (for satellites). 
The stellar mass is obtained by integrating the SFR along the branches of 
the halo merger tree. In what follows these galaxies 
are referred to as {\em model galaxies}.

\autoref{fig:model_mass} displays the stellar mass function for 
model galaxies in the simulation, plotted as solid circles connected
by the orange line. For comparison, the stellar mass function
estimated by \citet{Weaver2022gsmf} from the COSMOS galaxy sample is
plotted as solid squares connected by the green line. Note that 
this plot is not meant to make a quantitative comparison 
between our model prediction and observational data. Rather, 
the qualitative match between the data and the model demonstrates 
that our model galaxy population provides a realistic sample to 
construct mock catalogs.

\subsubsection{Luminosity assignment for model galaxies}
\label{sec:assignluminosity}

\begin{figure}
	\includegraphics[width=\columnwidth]{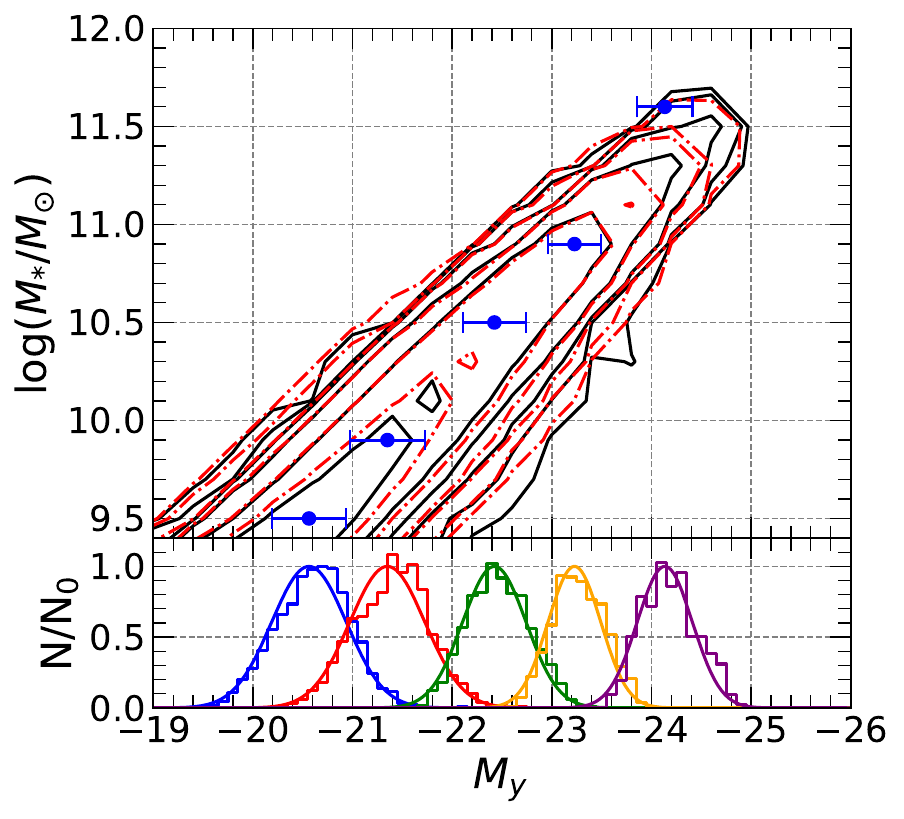}
    \caption{The upper panel shows the contours of the galaxy number 
    density in the plane of stellar mass $M_\ast$ versus $y$-band 
    absolute magnitude $M_{\rm y}$. The black lines 
    are for the data from COSMOS2020 and the red dash lines are for 
    model galaxies in our simulation. 
    The lower panel shows the histogram of $M_{\rm y}$ in different intervals 
    of $\log M_\ast$ and the lines are the best-fit Gaussians. 
    The blue dots with error bars in the upper panel show the 
    $M_\ast$-$M_{\rm y}$ relation and the Gaussian FWHM obtained from 
    the fits in the lower panel.}
    \label{fig:mass_luminosity}
\end{figure}

We assign luminosities in different bands to each of the model
galaxies. This is done in two steps. First we obtain the
luminosity in a given band, represented by the corresponding 
rest-frame absolute magnitude ($M_1$), according to the relationship
between the stellar mass and $M_1$ calibrated by observations (see below). 
Second, the luminosities in other bands are determined using the 
corresponding color indices, $M_1-M_2$, where $M_2$ is the 
rest-frame absolute magnitude in another band. In this step 
we also use the age of a sub-halo when determining the color of 
the galaxy it hosts, as detailed below. 

We use the photometric data in the COSMOS2020 catalog
\citep{Weaver2022} to calibrate our model for luminosities of 
model galaxies. The COSMOS field has been observed in 38 bands, 
covering a wide wavelength range from the ultraviolet to the far-infrared. 
The data thus enables precise estimates of both photometric redshifts 
(photo-z) and other galaxy properties, such as stellar mass and multi-band 
luminosities. We have estimated the galaxy stellar mass function using
the COSMOS galaxy sample provided by \citet{Weaver2022}. The result, 
plotted as the triangles connected by the blue line in 
\autoref{fig:model_mass}, is in
good agreement with the stellar mass function published in
\cite{Weaver2022gsmf}. We use the public software \verb'CIGALE'
\citep{Boquien2019} to fit the spectral energy distribution (SED) of
the COSMOS galaxies, and estimate the rest-frame absolute magnitude in
a given band by convolving the best-fit rest-frame spectrum with the
photometric response curve of the band. We adopt the photometric
redshifts provided by \citet{Weaver2022} during the fitting. 

\begin{figure}
  \includegraphics[width=\columnwidth]{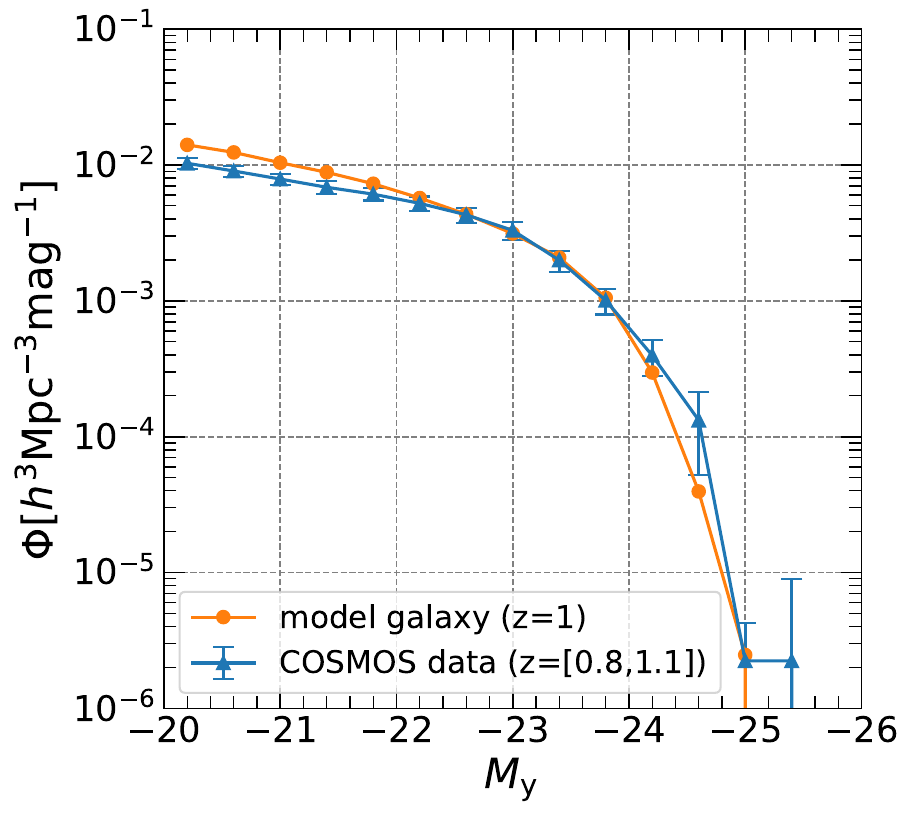}
  \caption{The $y$-band luminosity function of galaxies  
    at $z\sim1$. The orange curve is for the model galaxies in 
    the simulation and the blue curve is estimated from 
    the COSMOS2020 data which we use to calibrate our method  
    to assign the luminosity and colors for our model galaxies.}
  \label{fig:model_luminosity}
\end{figure}

We divide the galaxies in the COSMOS2020 catalog into successive
bins of stellar mass $M_{*}$, each with a fixed width of 0.2 dex.
For each $M_\ast$ bin, we fit the distribution of $M_{1}$
(absolute magnitude of the chosen band) with a Gaussian function.
For a model galaxy falling in the same $M_\ast$ bin,
we randomly assign an absolute magnitude according to the Gaussian fit.
\autoref{fig:mass_luminosity} shows the distribution of
the COSMOS2020 galaxies in the $\log M_\ast$ - $M_{\rm y}$ plane, 
where $M_{\rm y}$ is the $y$-band absolute magnitude. The black contours are
based on the observational data, while the red dashed contours are
based on the assigned $M_{\rm y}$. The lower panel shows the
$M_{\rm y}$ distributions and the corresponding Gaussian fits in 
different mass bins.
As can be seen, the assigned absolute magnitudes reproduce the
mass-luminosity relation very well, in terms of both the average
relation and the scatter. \autoref{fig:model_luminosity} compares
the $y$-band luminosity functions at $z\sim1$ estimated from the
COSMOS2020 sample using the real data (blue triangles) and the
assigned absolute magnitudes (orange circles). Again the two 
functions agree with each other very well.

\begin{figure}
  \includegraphics[width=\columnwidth]{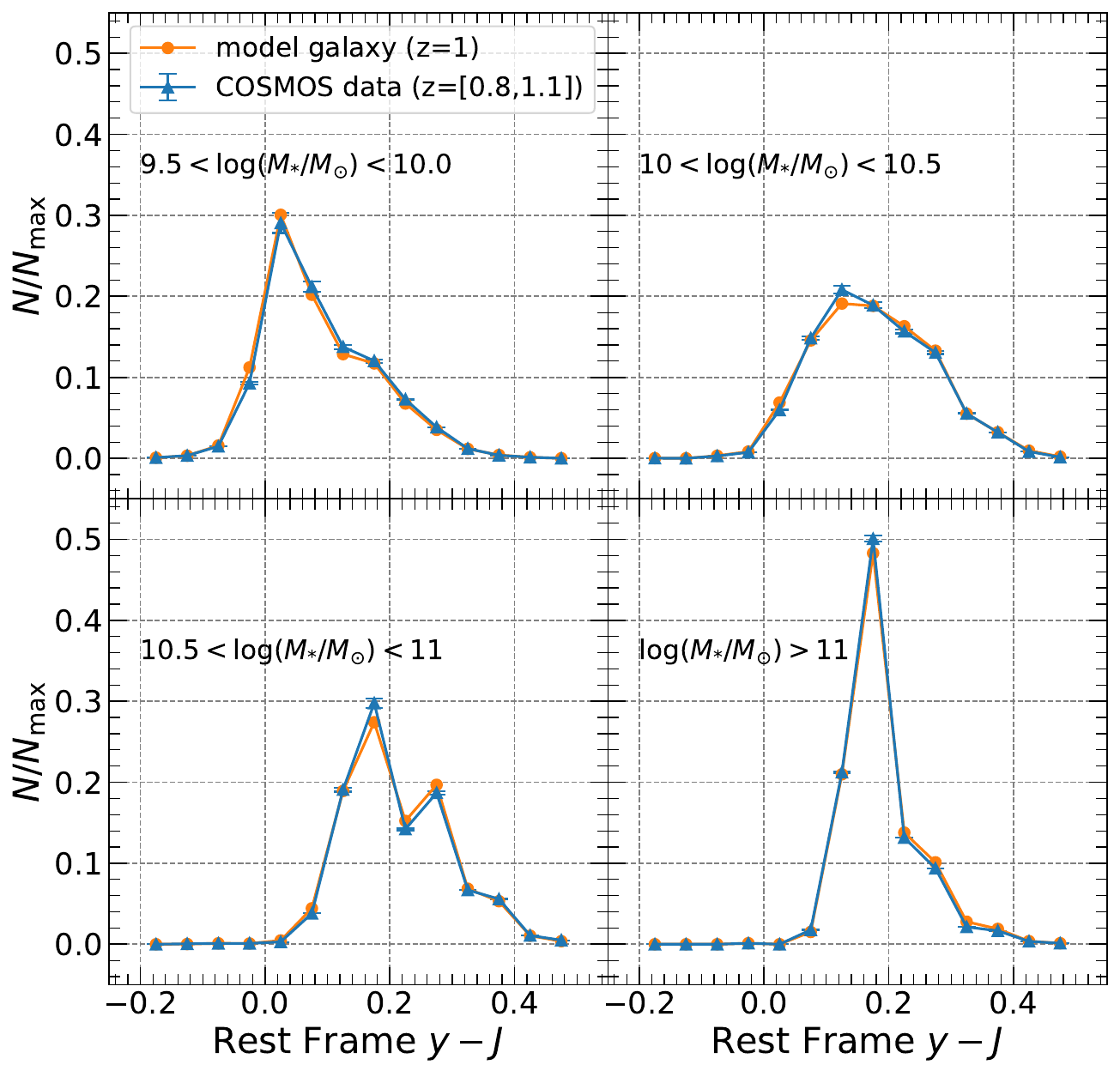}
  \caption{The distribution of the rest-frame $y-J$ color for 
  model galaxies (orange curves) and obtained from the COSMOS2020 
  data (blue curves with error bars).}
    \label{fig:model_color}
\end{figure}

We assign absolute magnitudes in bands other than $M_1$
by applying the sub-halo age distribution matching method
\citep{Hearin2013,Hearin2014,Wang2023}.  The method assumes that, at fixed 
stellar mass, the color of a galaxy is a monotonous function 
of the age of its host halo, usually quantified by $z_{\rm starve}$, 
the redshift at which the host halo started to be starved 
of cold gas supply. In general, $z_{\rm starve}$ is expected to be 
related to the halo mass assembly history. The following three definitions 
have been adopted for the same purpose \citep[e.g.][]{Wechsler2002,Behroozi2013,Hearin2013}:
\begin{enumerate}
\item $z_{\rm char}$, the highest redshift at which the
halo mass exceeds $10^{12}{\rm M}_{\odot}/h$;
\item $z_{\rm acc}$: the redshift at which a halo becomes a subhalo
  and remains such thereafter;
\item $z_{\rm form}$: the redshift after which the mass accretion 
  by the halo is negligible.
\end{enumerate}
Following common practice, we define $z_{\rm starve}$ as the 
maximum of the three redshifts defined above, 
\begin{equation}
  z_{\rm starve}={\rm Max}\{z_{\rm char},z_{\rm acc},z_{\rm form}\}.
\end{equation}

It is known that galaxies of given mass or luminosity can be
divided into two populations in the color space. In each stellar mass
interval, we fit the distribution of the color index $M_1-M_2$ ($M_2-M_1$ if the 2nd band is bluer) of the
COSMOS2020 galaxies by a bi-Gaussian function, so as to obtain  
a mass-dependent color distribution function, $n(M_1-M_2|M_\ast)$.
Similarly, for model galaxies, we obtain the mass-dependent 
distribution function of $z_{\mbox{starve}}$, $n(z_{\rm starve}|M_{*})$.
For each model galaxy, we then assign a $M_1-M_2$ color by matching
the number density of the observed galaxies above a given color 
threshold to that of the model galaxies above a certain 
threshold in $z_{\rm starve}$:
\begin{equation}
  n_{\rm mod}(>z_{\rm starve}|M_\ast)=n_{\rm obs}[>(M_1-M_2)|M_\ast].
\end{equation}
By solving this equation we obtain a monotonic relation between
$z_{\rm starve}$ and $(M_1-M_2)$, which enables us to assign an
$(M_1-M_2)$ color (thus $M_2$ for the known $M_1$) to each model galaxy
according to its $z_{\rm starve}$. \autoref{fig:model_color} 
compares the distributions of the $(y-J)$ color of model galaxies 
in different stellar mass bins (orange curves) to the corresponding   
distributions obtained from COSMOS2020 (blue curves). 
The two agree with each other very well, demonstrating that the 
properties of the galaxy population are well represented by
our model galaxies.  
We choose $(M_1, M_2)= (M_{\rm y}, M_{\rm J})$ for the PFS-like survey, 
$(M_1, M_2)= (M_{\rm I}, M_{\rm B})$ for the zCOSMOS survey, and  
$(M_1, M_2)=(M_{\rm i_{\rm AB}}, M_{\rm B})$ for the VIPERS survey.

\subsection{zCOSMOS, VIPERS and the PFS galaxy evolution survey}
\label{sec:survey_intro}

The zCOSMOS is a large redshift survey of galaxies
in the COSMOS field, carried out with the VIMOS spectrograph on the
8-meter ESO Very Large Telescope \citep[e.g.][]{Lilly2007}.
The survey consists of two components.  The first is zCOSMOS-bright, 
a magnitude-limited sample of about 20,000 
galaxies with $I_{\rm AB}<22.5$ and $0.1\lesssim z\lesssim 1.2$ 
that covers the whole 1.7 deg$^2$ COSMOS field.  
The other is zCOSMOS-deep, a sample consisting of about 10,000 
galaxies with $1.5\lesssim z\lesssim 3$ 
selected through color-selection criteria in the
central $\sim1$ deg$^2$ of the COSMOS field. 
Both samples are targeted to have a sampling rate of 
$\sim$70\%, but the actual sampling rate is 56\% for zCOSMOS-bright 
\citep{Knobel2012} and 55\% for zCOSMOS-deep \citep{Diener2013}. 
For our analysis, we only use the zCOSMOS-bright Sample.

The VIMOS Public Extragalactic Redshift Survey (VIPERS) explores 
a large cosmic volume in the redshift range $0.5<z<1.2$ \citep{Garilli2014,Scodeggio2018}.  Targets of the VIPERS were 
selected from the W1 and W4 fields of the Canada-France-Hawaii Telescope 
Legacy Survey Wide (CFHTLS-Wide) which covers a total sky area of 23.5 deg$^2$,
by applying an  apparent magnitude limit $i_{\rm AB}=22.5$, as well as 
a color selection on the $(r-i)$ vs. $(u-g)$ plane to remove  
galaxies at $z<0.5$. The selected targets were alaso observed with 
the VIMOS spectrograph on the ESO Very Large Telescope.
The complete sample contains 86,775 galaxies. The average target
sampling rate is about 47\% and the spectroscopic success rate is 
about 83\% \citep{Scodeggio2018}. The two rates combined give 
a total sampling rate of $\sim$40\%. 

The Prime Focus Spectrograph (PFS) project \citep{Takada2014} is one
of the next-generation multi-object spectroscopic surveys to be
accomplished on the Subaru 8.2-meter telescope. The PFS is a massively
multiplexed, optical and near-infrared (NIR) fiber-fed spectrometer,
equipped with 2394 re-configurable fibers distributed in a wide
hexagonal field of view with a diameter of 1.3 degrees. The PFS
project will conduct three major survey programs, dedicated to
fundamental and important questions in cosmology, galaxy evolution
and the origin of the Milky Way, respectively.  In this paper we
consider only the galaxy evolution survey, for which the science goals, 
sample selection and survey design are detailed in \cite{PFS-galaxysurvey}.
In particular, we will consider only the {\em Main sample} of the 
PFS galaxy evolution survey, which aims to obtain spectroscopy 
for $\sim 230,000$ galaxies down to stellar mass $M_\ast\approx3\times10^{10}M_\odot$ at $0.7<z<1.7$, 
with exposure times of 2 hours. Targets of the sample are selected 
to fall in the expected redshift range using photometric redshifts 
estimated from the available imaging data, and are limited by the 
$y$-band apparent magnitude as $y<22.5$. At $z>1$, targets 
fainter than $y=22.5$ may also be included if their $J$-band magnitude 
$J<22.8$. The redshift sampling rate is about 50\% for the sample 
at $z<1$, and 70\% at $z>1$. The sample galaxies are distributed 
over three separated fields (E-COSMOS, XMM-LSS and DEEP2-3) that 
have deep imaging in both optical and NIR. The total sky area of these 
fields is $\sim 12.3\,{\rm deg}^2$, and will be covered by 11 PFS pointings
(see Fig.8 in \citealt{PFS-galaxysurvey}).  Note that the analyses presented 
here use an earlier version of the survey design, which 
aimed at 13 PFS pointings with a total sky area of 
$\sim 14.5\,{\rm deg}^2$. Because of this (slight)
difference and the fact that the survey design may be further 
tuned, we will refer the survey as `PFS-like survey' in what follows. 

\subsection{Constructing mock catalogs}
\label{sec:mock_construction}

We construct mock catalogs for a given survey following the 
commonly-used two-step
method detailed in \citet{Blaizot2005}. First,   
an observing lightcone covering the same volume and redshift range as
the real survey is constructed using simulation snapshots in the same 
redshift range as the survey.  Next, model
galaxies are selected in the lightcone to form a mock catalog, and
their apparent properties, such as apparent magnitudes in different
bands, are computed taking into account selection effects  of 
the real survey. 

\subsubsection{Constructing lightcones}

Due to its limited box size, the simulation  box of a given snapshot 
has to be stacked to achieve a sufficiently large volume,  
taking advantage of the periodic boundary condition. 
However,  because of the periodic boundary condition the same 
structure  can appear
repeatedly if observed along a coordinate axis of the simulation box.
Following \citet{Blaizot2005}, we apply the technique of random tiling
to overcome this problem. 
By applying  this random tilling  scheme many times, we can generate 
a set of different mock catalogs, in which the same (simulated) universe 
is virtually observed from different directions.

The stacked and randomly transformed snapshots of different
redshifts are then used to construct the lightcone. We fill
in successive intervals of comoving distance with the corresponding 
snapshots. Specifically, for an interval of comoving distance, 
\begin{equation}
	\frac{D_{i}+D_{i-1}}{2}<D<\frac{D_{i}+D_{i+1}}{2},
\end{equation}
we use model galaxies from the snapshots at redshift $z_{i}$.
For each model galaxy, we calculate a cosmological redshift 
$z_{\rm cos}$ from its comoving distance. The peculiar
velocity along the line of sight, $v_{\rm pec}$, 
which is estimated using the velocity of the (sub-)halo 
which the galaxy live in, is added
to the $z_{\rm cos}$ to give a spectroscopic
redshift $z_{\rm spec}$:
\begin{equation}
  z_{\rm spec}=\sqrt{\frac{1+v_{\rm pec}/c}{1-v_{\rm pec}/c}}(1+z_{\rm cos})-1.
\end{equation}
We also estimate a photometric redshift (photo-$z$) for each galaxy
by including the typical uncertainty of photo-$z$ 
of the survey. Assuming the photo-$z$ error in $\frac{\Delta z}{1+z}$
to be $\sigma$, we have 
\begin{equation}
  z_{\rm photo}=z_{\rm spec}+(1+z_{\rm spec})\times N(0,\sigma^{2}),
\end{equation}
where $N(0,\sigma^{2})$ is the normal distribution with the mean 
equal to zero and with the standard deviation equal to $\sigma$.

\begin{figure*}
  \includegraphics[scale=0.35]{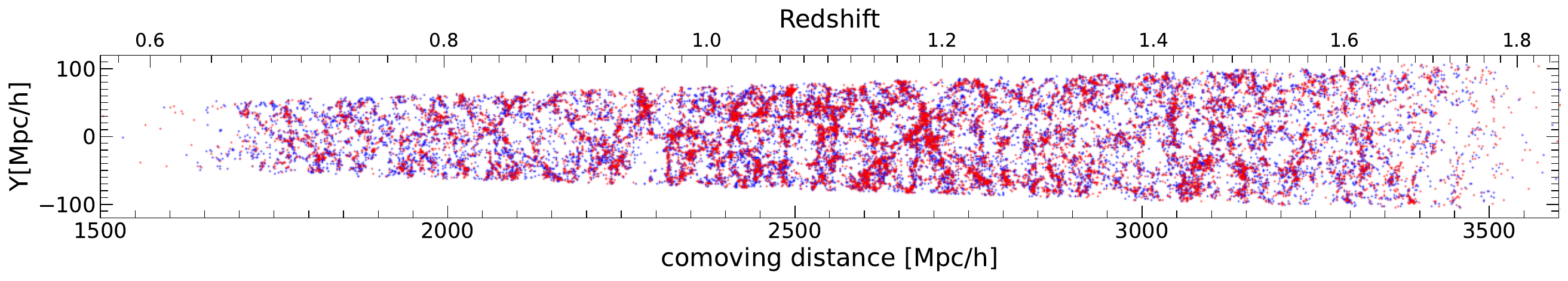}
  \caption{The lightcone of the PFS-like survey in the XMM-LSS field. 
  The range of the declination 
  is $-4.9^{\circ} \le \delta \le -4.4^{\circ}$. The color of a point represents the 
  intrinsic $(y-J)$ color of the galaxy.}
  \label{fig:lightcone}
\end{figure*}

Finally, we finish the construction of the lightcone by excluding 
model galaxies outside the sky coverage of the survey.  As mentioned
above, by applying the random tilling many times, we can generate a
set of different lightcones to form a set of mock catalogs.
As an example, \autoref{fig:lightcone} shows
one of the lightcones made for one field of the PFS-like survey, 
projected onto the RA and redshift plane. 
The red and blue dots represent galaxies with red and blue colors 
according to $(y-J)$, respectively.

\subsubsection{Incorporating observational selection effects}
\label{sec:mock}

We calculate apparent magnitudes of model galaxies in a given
lightcone according to their absolute magnitudes and 
redshifts, and we construct a mock catalog by applying the same
selection effects as the real survey. The apparent magnitude of the
$i$-th band is given by 
\begin{equation}
  m_i=M_i+5\log_{10}(D_{L})+25+K\cdot \log_{10}(1+z),
\end{equation}
where $M_i$ is the absolute magnitude of the $i$-th band ($i=1,2$),
$D_L$ is the luminosity distance in units of Mpc, and $K$ is the
$k$-correction. The value of $K$ is not available from our model.
We estimate the value of $K$ using the COSMOS2020 galaxy catalog,
assuming that galaxies of the same luminosities and colors
have a similar spectral energy distribution. For a given model
galaxy, we identify a real galaxy from the COSMOS2020 catalog
by matching the absolute magnitude, $M_1$, and the color index
$M_1-M_2$, where the bands for $M_1$ and $M_2$ may be different
for different surveys (e.g. $M_{\rm y}$ and $M_{\rm J}$ 
for the PFS galaxy survey). The $K$ value of the real galaxy,
which was estimated from the best-fit rest-frame spectrum of
the galaxy by spectral fitting (see \S~\ref{sec:assignluminosity}), 
is then assigned to the model galaxy. 

The details of the target selection strategy may vary from survey to
survey, but in general targets are selected to form a
magnitude-limited sample according to the apparent magnitude in a
specific band. For instance, the SDSS main sample targeted galaxies
with $r$-band magnitude $r\lesssim17.77$ \citep{York-00}. For high-$z$
surveys, additional redshift criteria, based on  
color-color diagrams or photometric redshifts, 
are used to exclude targets outside the aimed redshift range. 
The PFS-like survey will observe galaxies with $0.7<z<1.7$ and 
$y<22.5$ (if at $0.7<z<1.7$) or $J<22.8$ (if at $z>1$). 
The zCOSMOS-bright sample selects galaxies with $I<22.5$.
The VIPERS uses the magnitude limit of $i_{\rm AB}<22.5$.

In most spectroscopic surveys, especially those at high redshift, only
a fraction of the targets satisfying the redshift and magnitude limits
can be included in the spectroscopic sample,  due to the limited
number of fibers, finite observing time,  imperfect redshift
identification and a variety of other limitations (e.g. obscuration by
bright stars, bad weather, fiber collisions). In general, these
effects combined can be quantified by a sampling rate, 
defined as the fraction of the targets that are actually observed and 
included in the final sample. The sampling rate may vary across the 
survey coverage as well. 
For the PFS-like survey, the angular sampling is expected to vary
across the survey area. The inhomogeneity is mainly caused by
the fact that the spectroscopic target sampling becomes increasingly
low in the sky where the target density is high, because two fibers
cannot be positioned too closely.  The fiber assignment strategy of the
PFS survey is described in \cite{Sunayama2019} and 
\citet{Shimono2016}, and a software named
Exposure Targeting Software (ETS)
\footnote{\href{https://github.com/Subaru-PFS/ets_fiberalloc}
{https://github.com/Subaru-PFS/ets\_fiberalloc}.}
is designed to implement the assignment for a given sample of
targets. We apply this software to our mock catalogs, by requiring an
average sampling rate of 50\% for $0.7<z<1$ and 70\% for $1<z<1.7$, 
as planned for the upcoming PFS galaxy survey. 

For the zCOSMOS-bright sample, the RA-dependence of the sampling rate 
due to incomplete targeting was presented in figure 5 of \citet{Torre2011}, while 
the redshift-dependent sampling due to imperfect redshift identification (redshift 
success rate) was shown in figure 9 of \citet{Lilly2007}. The spatial variation of 
the target sampling rate and the redshift measurement success rate for the VIPERS 
survey were described in figure 5 and figure 6 of \citet{Scodeggio2018}. 
In addition, the VIPERS survey has a redshift-dependent sampling rate caused 
by the color selection used to target galaxies at $z>0.5$. The sampling rate as 
a function of redshift was given in figure 4 of \citet{Scodeggio2018}. 

By incorporating these selection effects, we have constructed a set of 20 
mock catalogs for each of the three surveys. For the zCOSMOS-bright 
survey and the VIPERS, our mock catalogs respectively contain 
$15,611\pm 1,014$ and $77,927\pm 1,081$ galaxies, in good agreement 
with the sample sizes of the real surveys: 16,604 for zCOSMOS-bright 
and 75,765 for VIPERS. 

\section{Effects of flux limit criteria on galaxy clustering}
\label{sec:flux_limit_effect}

As mentioned above, statistics of the galaxy distribution 
are mainly affected by three incompleteness effects, i.e. 
sampling rate effect, fiber collision effect and 
flux limit effect. The former two effects have been 
discussed and understood in depth in the literature based on 
studies of previous surveys. In contrast, the {\em flux limit effect}, 
as initially noticed by \citet{Meneux2008,Meneux2009}, has been largely ignored. 
In this section, we focus on this effect.
We propose a new method to correct it and use mock catalogs to test  
the method. 

\subsection{Effects of flux-limit criteria}
\label{sec:effect_description}

\begin{figure*}
  \includegraphics[width=\textwidth]{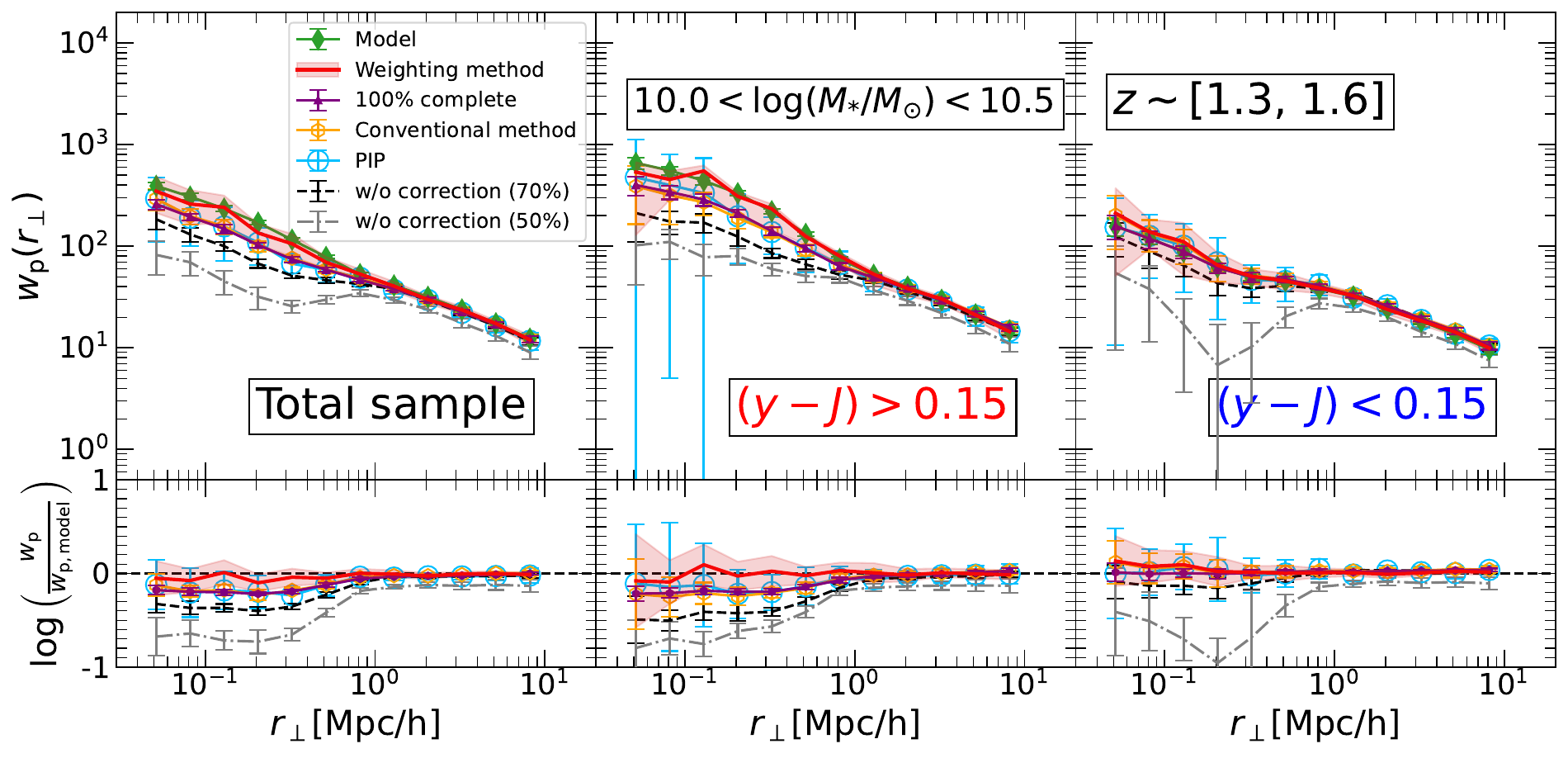}
  \caption{Measurements of $w_{\rm p}(r_{\perp})$ for the total sample 
  (left panels) and subset of red galaxies (middle panels) 
  and blue galaxies (right panels), using the PFS-like 
  mock catalogs with stellar mass limited to 
  $10<\log(M_\ast/{\rm M}_\odot)<10.5$ between redshift 1.3$\sim$1.6. The different lines and symbols represent different methods as indicated in the figure. }
\label{fig:wp_cor}
\end{figure*}

The flux limit effect may cause an underestimated 
measurement of the small-scale galaxy clustering, and 
the effect is more significant for red galaxies with low 
stellar mass compared to blue galaxies of similar masses. 
The reason is that, at fixed mass, red galaxies in general 
have larger mass-to-light ratios than blue galaxies. Thus, 
for a sample of galaxies in a given stellar mass range, red 
galaxies have a higher probability to be excluded by the 
flux limit. This effect is strong particularly for 
low-mass red galaxies which are predominately satellite 
galaxies in massive halos \citep[e.g.][]{Li-06c,Li-07,LanMenardMo2016},
and so are more strongly clustered than their blue 
counterparts \citep[e.g.][]{Li-06b,Li-06c,Li-07,Zehavi_etal2011}. 
Therefore, the reduction of this population 
in the sample can lead to an underestimate of the one-halo term 
of the correlation function on scales comparable to and smaller 
than the size of dark matter halos.  The flux limit effect is expected 
to be more important at higher $z$ and when the observational 
band used for target selection corresponds to a bluer rest-frame band. 

In order to highlight the flux limit effect, 
we have constructed a set of ``100\% complete''
mock catalogs that are exactly the same as the PFS-like mock 
sample except that the sampling rate is set to 100\%. 
As can be seen from \autoref{fig:wp_cor}, the projected 
two-point correlation function $w_{\rm p}(r_{\perp})$ of 
the complete (100\%) sample (purple triangles/lines) so obtained is 
significantly lower than the true $w_{\rm p}(r_{\perp})$ 
(green diamonds/lines) at scales smaller than a few Mpc. In the figure we show the comparison
only for the lowest mass bin with $10<\log (M_\ast/{\rm M}_\odot)<10.5$
and the relatively high redshift bin with $1.3<z<1.6$, where 
we find the effect is the most pronounced. In the figure we also 
show the same comparison but for red and blue galaxies separately. 
Clearly, the underestimation of the small-scale clustering seen 
in the total sample is dominated by red galaxies, as expected.
For the mass and redshift bins shown, the small-scale clustering is underestimated by 
about 0.4 dex for the total sample, and about 0.5 dex and 0.1 dex 
for the red and blue subsamples, respectively. 
In the rest of this section we attempt to develop a method to 
correct for the flux limit effect.

\subsection{Conventional methods of measuring galaxy clustering}
\label{sec:conventional_method}




Conventionally, a number of estimators have been used to measure
the correlation function \citep[][]{Davis1983,Hamilton1993,Landy1993}. 
Here we use the Landy-Szalay estimator \citep{Landy1993}. 
For a given galaxy sample {\tt D} and the corresponding 
random sample {\tt R} that is constructed to have the same selection 
effects as sample {\tt D} (more details can be found in \citealt{Li-06a}), 
the redshift-space correlation function is estimated using 
\begin{equation}
    \label{eq:LS}
  \xi (r_{\perp},\pi)=\frac{1}{RR}\times 
  \left[\frac{N_{r}(N_{r}-1)}{N_{d}(N_{d}-1)} DD-2\frac{N_{r}}{N_{d}} DR+RR\right],
\end{equation}
which is a function of the pair separations both perpendicular 
($r_{\perp}$) and parallel ($\pi$) to the line of sight. Here, 
$N_{\rm d}$ and $N_{\rm r}$ are the sample sizes of {\tt D} and {\tt R}, 
respectively, and $DD$, $RR$ and $DR$ are the counts of galaxy-galaxy 
pairs in the galaxy sample, random-random pairs in the random 
sample, and galaxy-random cross pairs between the two samples, 
respectively.  The pair counts are also functions of $r_{\perp}$ and $\pi$.
To reduce effects of redshift distortions, we use the projected 2PCF,
defined as the integration of $\xi(r_{\perp},\pi)$ along the line of sight:
\begin{equation}
  w_{\rm p}(r_{\perp})=2\int_{0}^{+\infty}\xi(r_{\perp},\pi){\rm d}\pi
  =2\sum_{i=1}^{n} \xi(r_{\perp},\pi_{i})\Delta\pi_i.
\end{equation}
We choose $n=40$ and $\Delta\pi_i=$ 1 ${\rm Mpc/h}$, 
so that the summation runs from $\pi_1=$ 0.5 ${\rm Mpc/h}$ up to 
$\pi_{40}=$ 39.5 ${\rm Mpc/h}$ with a constant interval of 1 ${\rm Mpc/h}$.

In general, the incompleteness of sample {\tt D} may not be fully 
incorporated in the random sample {\tt R}, and some weighting scheme 
is needed to correct for the residual 
incompleteness. Here we consider two weighting schemes that have been
used in the literature. In the first one, effects of sampling rate,
redshift success rate and fiber collision are considered separately, 
and a final weight is given to each galaxy by combining all the effects. 
To this end, we use the Voronoi tessellation to estimate the sky 
position-dependent sampling rate, $f_{\rm s}({\rm RA},{\rm Dec})$. 
The tessellation divides the whole survey area into non-overlapping 
polygons, so that each polygon contains one (and only one) observed 
galaxy and that every position within the survey area is covered by 
one (and only one) polygon. The sampling rate of the $J$-th polygon 
is estimated by the inverse of the total number of targets in 
the parent photometric sample: $f_{{\rm s},J}=1/N_{{\rm phot},J}$. 
Thus, the sampling rate is a constant within a 
given polygon, and each galaxy in the observed sample 
is assigned the sampling rate of the polygon in which it 
resides: $f_{{\rm s},i}=f_{{\rm s},J}$, where 
$i$ refers to the $i$-th galaxy and $J$ to the $J$-th polygon. 

The redshift success rate for each galaxy 
in the spectroscopic sample, $f_{\rm z}(z,{\rm \theta_k})$, 
is a function of redshift, $z$, and a set of 
properties of the galaxy, ${\rm \theta_k}$.
In principle, one could consider $n$ properties, and estimate 
the success rate function, $f_{\rm z}(z,{\rm \theta_k})$, 
in the $(n+1)$-dimensional parameter space using the ratio 
between the number of successfully-measured 
redshifts, $N_{\rm succ}$, and the number of spectroscopic targets, 
$N_{\rm obs}$. One can then estimate $f_{{\rm z},i}$ for each galaxy 
in the sample.
Once $f_{{\rm s},i}$ and $f_{{\rm z},i}$ are obtained 
for the $i$-th galaxy, we assign to it a weight defined as  
\begin{equation}
\label{eq:w_sky}  
w_{{\rm sky},i}=\frac{1}{f_{{\rm s},i}f_{{\rm z},i}}. 
\end{equation}
This weight will be used to correct for the effect of 
sampling rate and redshift success rate of the survey when we estimate 
both the number density and clustering of galaxies. 



Following \cite{Hawkins2003} and \cite{Li-06b}, we correct
for the effect of fiber collisions using angular correlation function of 
the parent photometric sample. We first estimate the angular correlation 
for both the parent photometric sample
($w_{\rm p}(\theta)$) and the spectroscopic sample 
($w_{\rm s}(\theta)$). Note that $w_{\rm s}$ is corrected 
for the sampling rate and redshift success effect as described above. 
When computing the galaxy-galaxy pair 
count $DD$ in \autoref{eq:LS}, we weight each pair by 
\begin{equation}
  \label{eq:f_theta}
  w_{{\rm fiber}, ij}(\theta_{ij})=\frac{1+w_{\rm p}(\theta_{ij})}{1+w_{\rm s}(\theta_{ij})},
\end{equation}
where $\theta_{ij}$ is the angular separation of the pair between the $i$-th galaxy and the $j$-th galaxy. 
As demonstrated in \citet{Li-06a}, this weighting method 
works well in correcting the underestimation of the clustering 
on scales where fiber collision effect is significant. The weighting scheme works as follow: for the pair formed by the $i$-th and $j$-th galaxy, we compute their separations perpendicular ($r_{\perp, ij}$) and parallel ($\pi_{ij}$) to the line-of-sight, and the angular separation ($\theta_{ij}$). We only use $\theta_{ij}$ to calculate the weight $w_{{\rm fiber},ij}$ for the pair formed by the $i$-th and $j$-th galaxy using \autoref{eq:f_theta}. Then we count the galaxy-galaxy pairs as function of $r_{\perp}$ and $\pi$. The counts of galaxy-galaxy pairs are computed 
using $DD(r_{\perp},\pi)=\sum w_{{\rm sky},i}w_{{\rm sky},j}w_{{\rm fiber},ij}$, where the sum runs over all the galaxy-galaxy pairs 
of given separation ($r_{\perp}$, $\pi$) in the observed sample. The counts of galaxy-random pairs are 
computed using $DR(r_{\perp},\pi)=\sum w_{{\rm sky},i}$, where the sum runs over all galaxy-random pairs of given separation 
($r_{\perp}$,$\pi$) between the observed and random samples.

The weighting scheme described above has been widely adopted in 
previous studies of galaxy clustering, and is labelled as the ``conventional method''
in what follows. Applying this method to our mock catalogs,
we obtain the $w_{\rm p}(r_{\perp})$ measurements and plot them as orange 
circles/lines in \autoref{fig:wp_cor}, for the same mass 
and redshift bins. The black dashed lines represent the results 
without any corrections where the sampling rate is about 70\%. 
We also plot results with a sampling rate 50\% as the gray dash-dot lines. It can be 
seen that even $w_{\rm p}(r_{\perp})$ on large scales can be affected by incomplete sampling. As can be seen, the conventional method 
fails to reproduce the {\em true} correlation function, 
although it successfully reproduces the correlation function 
of the ``100\% complete'' sample. 

We also consider another weighting scheme originally proposed 
by \citet[][]{Bianchi2017} and referred to as the
pairwise-inverse-probability (PIP) scheme in the following. 
As shown in \citet[][]{Bianchi2017} and \citet{Bianchi2018},
this weighting scheme is able to correct the {\em sampling rate effect} 
(including redshift success rate) and {\em fiber collision effect} 
simultaneously. In short, for a given spectroscopic galaxy 
redshift survey and its parent photometric sample, the target 
selection software that is used for selecting the spectroscopic 
targets is rerun $N$ times, each with a different 
random number seed. This gives rise to $N$ realizations of 
the target selection process, resulting in a set of target 
samples with sizes equal to the real sample. Suppose that the 
galaxy-galaxy pair formed by the $i$-th and $j$-th galaxy is 
selected $b$ times during the $N$ realizations, the pair is 
given a weight of $w_{ij}=N/b$. If the $i$-th galaxy is selected 
$c$ times in total, the galaxy gets a weight of $w_{i}=N/c$. 
The PIP weights $w_{ij}$ are then used to weight galaxy-galaxy 
pairs $DD$, while the individual weights $w_i$ are used to weight 
the galaxy-random cross pairs $DR$ (see equation
13 in \citealt{Bianchi2017}). The galaxy-galaxy pair count is then given by 
\begin{equation}
  DD(r_{\perp}, \pi)=\sum w_{ij}\frac{DD^{\rm (p)}_{a}(\theta_{ij})}{DD_{a}(\theta_{ij})}, 
\end{equation}
where $DD^{\rm (p)}_{a}(\theta)$ and $DD_{a}(\theta)$ represent 
the angular pair counts of the parent sample and the observed sample, 
respectively. For the pair formed by the $i$-th and $j$-th galaxy, the angular separation $\theta_{ij}$ is only used to compute $DD^{\rm (p)}_{a}(\theta)$ and $DD_{a}(\theta)$. The sum runs over all the galaxy-galaxy pairs 
of given separation ($r_{\perp}$, $\pi$) in the observed sample. 
$DD^{(p)}_{a}(\theta)$ is estimated from the parent sample, while 
$DD_{a}(\theta)$ is computed from the observed sample 
using the weights $w_{ij}$, i.e. $DD_{a}(\theta)=\sum w_{ij}$. 
Similarly, the galaxy-random pair count is given by 
\begin{equation}
  DR(r_{\perp}, \pi)=\sum w_{i}\frac{DR^{\rm (p)}_{a}(\theta)}{DR_{a}(\theta)}. 
\end{equation}
For the pair formed by the $i$-th galaxy and random point, the angular separation is only used to compute $DR^{\rm (p)}_{a}(\theta)$ and $DR_{a}(\theta)$. The sum runs over all galaxy-random pairs of given separation 
($r_{\perp}$,$\pi$) between the observed and random samples. 
$DR_{a}(\theta)$ is computed with the weights $w_{i}$, by 
$DR_{a}(\theta)=\sum w_{i}$. \citet{Bianchi2018} tested this 
weighting scheme using mock samples of the DESI survey and found 
that it can correctly deal with effects of fiber assignment 
({\em sampling rate effect} and {\em fiber collision effect}). 
We also test it here, by applying the above weights to galaxy-galaxy 
pairs and galaxy-random pairs in our mock catalogs. The resulting 
$w_{\rm p}(r_{\perp})$ are shown in \autoref{fig:wp_cor} as blue circles. 
We can see that the PIP method works equally well as the ``conventional 
method'' in reproducing the $w_{\rm p}(r_{\perp})$ of 
the ``100\% complete'' sample, but both methods cannot reproduce 
the {\em true} correlation function. In other words, the effect of 
flux limit is not taken into account by either of the two weighting schemes. 

Before moving on to the development of a new method, we point out 
that the conventional methods are still useful in certain cases,
e.g., on scales larger than a few Mpc and for samples with 
relatively high masses or low redshifts. As shown above, 
when applicable, the ``conventional method'' and the PIP method are 
essentially equivalent. Both methods rely on the parent photometric
sample. The PIP method has the advantage of 
correcting all effects (except the flux limit effect) 
simultaneously, but it is necessary to have the target selection 
software which is not always publicly available. In this regard, 
the ``conventional method'' method may be more applicable, although 
it assumes that different effects can be modeled separately using  
independent weights.

\subsection{A new method to correct for the flux limit effect}
\label{sec:new_method}

Here we propose a new method to correct the bias in the clustering 
measurements as caused by the flux limit effect. For this purpose
we use two photometric samples: one is the parent photometric 
sample ({\tt sample p}) that includes all galaxies in the same redshift and 
mass ranges as the redshift survey sample ({\tt sample s}), 
and the other is the subset of galaxies in {\tt sample p}
that meet the flux criteria of the survey sample ({\tt sample p$^\prime$}). 
We estimate the sampling rate caused by the flux limit selection criteria in 
the space spanned by galaxy properties $\vec{\mathbf{q}}$. Here we use 
the redshift, stellar mass and rest-frame color of the galaxies, that is, $\vec{\mathbf{q}}=\{z, M_\ast, y-J\}$. The redshift used here is the photometric redshift, and we assume that the photometric redshift, stellar mass and rest-frame color for galaxies in the photometric parent sample 
can be estimated from fitting the multi-band SED, e.g. 
by using data from the HSC imaging survey for the PFS targets. 
The sampling rate caused by the flux limit selection criteria is estimated 
directly by 
\begin{equation}
  \label{eq:sr_flux}  
  f_{\rm flux}\left(\vec{\mathbf{q}}\right)=\frac{N_{\rm p^\prime}\left(\vec{\mathbf{q}}\right)}{N_{\rm p}\left(\vec{\mathbf{q}}\right)}, 
\end{equation}
where $N_{\rm p^\prime}$ and $N_{\rm p}$ are the number of galaxies in {\tt sample p$^\prime$} and {\tt sample p}, respectively. For each galaxy in {\tt sample s}, we assign the weight defined as 
\begin{equation}
  \label{eq:w_flux}  
  w_{{\rm flux},i}=\frac{1}{f_{\rm flux}\left(\vec{\mathbf{q}}_{i}\right)} 
\end{equation}
to it when we count $DD$ and $DR$ in \autoref{eq:LS}. In our new weighting scheme, the counts of galaxy-galaxy pairs are computed using $DD=\sum w_{{\rm sky},i}w_{{\rm sky},j}w_{{\rm fiber},ij}w_{{\rm flux},i}w_{{\rm flux},j}$, and the counts of galaxy-random pairs are computed using $DR=\sum w_{{\rm sky},i}w_{{\rm flux},i}$. We notice that $f_{\rm flux}$ 
could be very small or even zero  in some regions of the space spanned by $\vec{\mathbf{q}}$. 
A small $f_{\rm flux}$ will give a very large weight to the galaxy and cause 
large shot noise in the measured $w_{\rm p}(r_{\perp})$, while a $f_{\rm flux}$ 
of zero will lead to an invalid weighting to the galaxy. To avoid this problem, 
we make a sampling rate cut in {\tt sample s}, excluding all galaxies 
from {\tt sample s} with $f_{\rm flux}\left(\vec{\mathbf{q}}\right)<2\%$.
We make the same cut for model galaxies in the original simulation box
in order for a fair comparison of galaxy clustering between the model 
and mock catalogs. 

We apply this method to our mock catalogs and plot the results
of $w_{\rm p}(r_{\perp})$ as solid red lines surround by 
pink bands in \autoref{fig:wp_cor}, as labelled by ``weighting method''. 
It is encouraging that this method reproduces the true 
$w_{\rm p}(r_{\perp})$ of the input model on all scales and for both red 
and blue galaxies. As we will show further 
in the next section, this method provides an unbiased measurement of 
the projected 2PCF for all the samples selected by a flux limit criterion 
as in PFS-like surveys. As demonstrated above, the flux-limit 
criteria can cause different sampling rates for galaxies of different properties 
(e.g. color). Our weighing method accounts for this sampling effect 
as well as the sampling effect caused by fiber assignments. However, 
the conventional method can only account for the latter.

\section{Number density and clustering of galaxies in PFS-like surveys}
\label{sec:mock_test}

In this section, we measure the abundance and clustering 
of galaxies in the mock catalogs we construct 
in \S~\ref{sec:mock_construction} for the PFS-like galaxy survey. 
The purpose of our analysis here is two-fold. First, we use the 
mock catalogs to test our methods of measuring galaxy number density 
and clustering, especially the new method to correct for the bias 
caused by the flux-limited selection criteria. 
Second, we use the mock catalogs to estimate errors in 
the measurements for different surveys, and we make comparisons 
between the PFS-like sample and the zCOSMOS and VIPERS samples. 
This allows us to predict the improvements expected from the upcoming PFS 
survey relative to the previous surveys. 

\subsection{Galaxy luminosity functions and stellar mass functions}
\label{sec:abundance}

\begin{figure*}
	\includegraphics[width=0.85\textwidth]{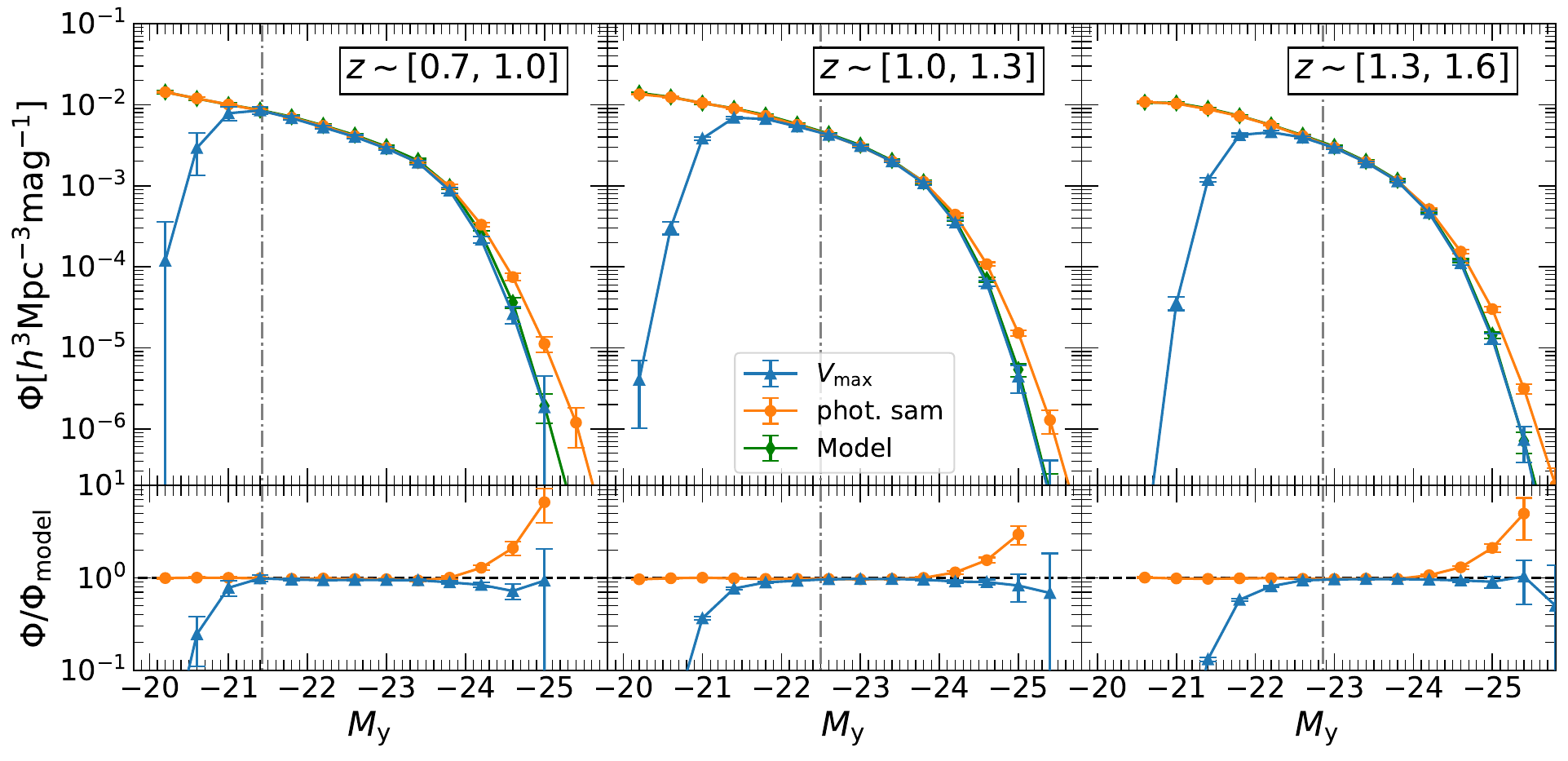}
    \caption{The $y$-band luminosity functions 
    estimated from the mock catalogs of the PFS-like galaxy survey 
    are plotted as blue dots with $V_{\rm max}$-weighting method and orange dots with photometric sample, and are compared to the 
    {\em true} luminosity function of model galaxies in the 
    simulation. Results are shown for three different redshift 
    intervals as indicated. The black vertical 
    line in each panel indicates the spectroscopic sample limit.}
    \label{fig:pfs_lf_y}
\end{figure*}

\begin{figure*}
	\includegraphics[width=0.85\textwidth]{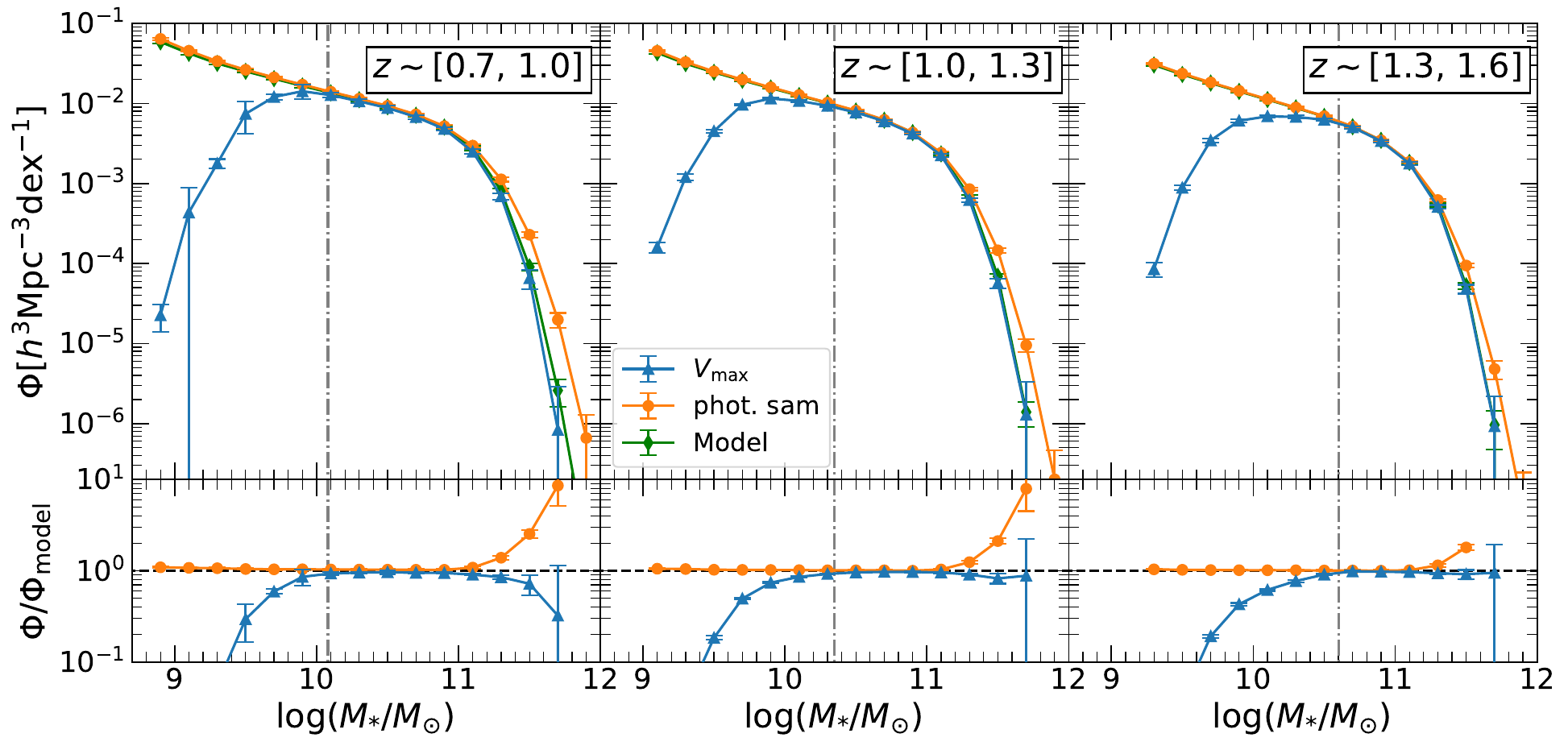}
    \caption{The galaxy stellar mass functions 
    estimated from the mock catalogs of the PFS-like galaxy survey
    are plotted as blue dots with $V_{\rm max}$-weighting method and orange dots with photometric sample, and are compared to the 
    true stellar mass function of model galaxies in the simulation.
    Results are shown for three different redshift intervals 
    as indicated. The black vertical 
    line indicates the spectroscopic sample limit.}
    \label{fig:pfs_gsmf}
\end{figure*}

\begin{figure*}
	\includegraphics[width=0.85\textwidth]{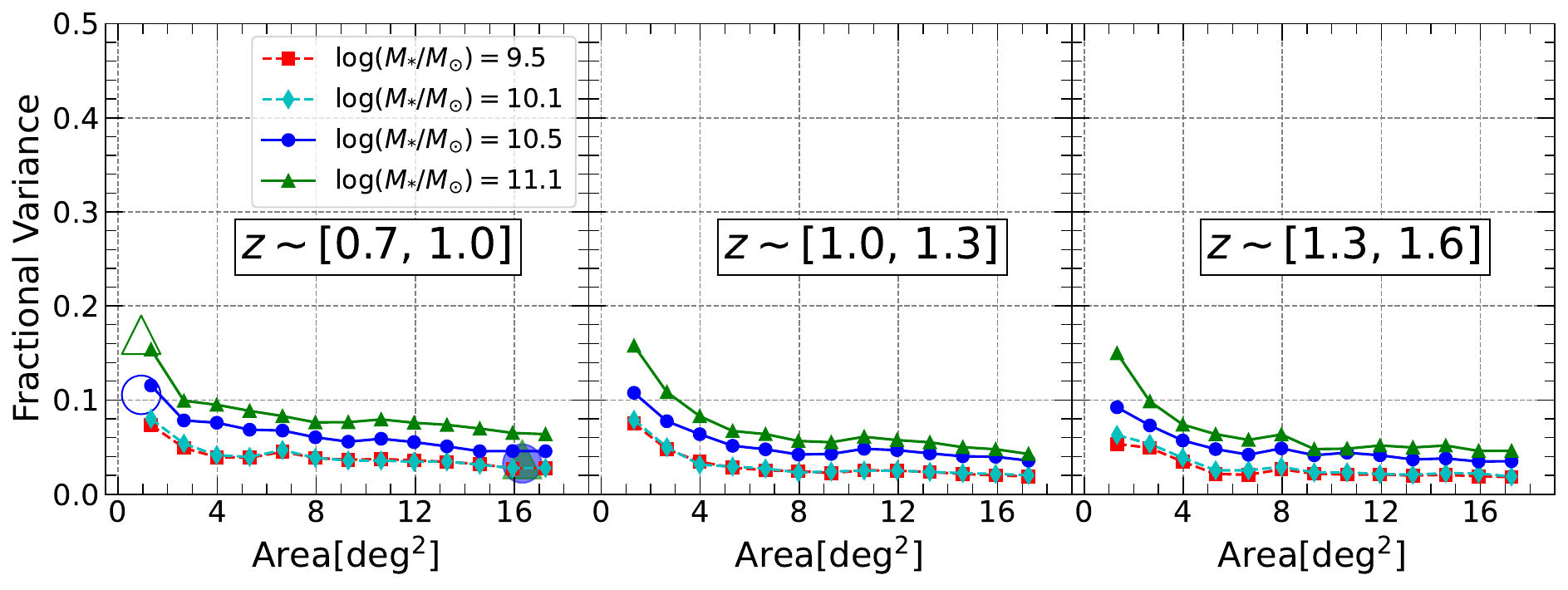}
	\caption{The fractional variance in the GSMF is estimated from the mock catalogs of the PFS-like galaxy survey, 
		as a function of the survey area. 
		Results are show for different redshift intervals in 
		different panels, and in each panel the different symbols/colors 
		are for different stellar masses as indicated. The solid lines represent the results with $V_{\rm max}$-weighting method from spectroscopic sample and the dashed lines represent the results from photometric sample. The 
		large hollow symbols are the fractional variance estimated 
		for the zCOSMOS-bright survey, and the large solid symbols are the results for VIPERS survey.}
	\label{fig:pfs_cv_smf}
\end{figure*}

We use the commonly-adopted $V_{\rm max}$-weighting method 
\citep{Schmidt1968} to estimate the luminosity functions (GLF) 
and stellar mass functions (GSMF) of galaxies in the mock catalogs
of the three surveys.  
For the $i$-th galaxy with an absolute magnitude 
$M_{1,i}$ in a given survey, we determine a maximum redshift, $z_{{\rm max},i}$, 
at which $M_{1,i}$ corresponds to the limiting apparent 
magnitude that was used to select the sample. 
Given the survey area, $z_{{\rm max},i}$ defines
a maximum volume, $V_{{\rm max},i}$, over 
which galaxy targets with the same absolute magnitude, 
$M_{1,i}$, can be observed in the apparent 
magnitude-limited sample. The inverse ratio of this 
volume to the total survey volume 
(given by the upper boundary of the redshift of the survey),
$V_{\rm survey}/V_{{\rm max},i}$, is used to weight the galaxy 
and to statistically correct for the luminosity-dependent 
incompleteness of the survey volume. 
The sample rate effect is corrected by multiplying this weight 
by the sky position-dependent weight $w_{{\rm sky},i}$ given by 
\autoref{eq:w_sky}. The total weight for a given galaxy is 
\begin{equation}
  \label{eq:w_tot}
     w_{{\rm tot},i} = \frac{V_{\rm survey}}{V_{\rm max}} \times w_{{\rm sky},i}
                     = \frac{V_{\rm survey}}{V_{\rm max}} \times \frac{1}{f_{{\rm s},i}f_{{\rm z},i}}.
  \end{equation}
Each galaxy in a mock catalog is then weighted by the total weight $w_{{\rm tot},i}$ when 
measuring the GLF and GSMF. 
Due to the limited depth of a survey, both the luminosity 
and stellar mass functions can only be measured down to a limited 
luminosity or mass. The limit for the luminosity function 
measured for galaxies at a given redshift can be determined by 
the apparent magnitude limit of the survey. 
For the stellar mass function, however, the determination
of the stellar mass limit $M_{*,{\rm min}}(z)$ is not straightforward. 
If the survey uses the apparent magnitude of the $j$th-band 
to select the targets, we use the symbol $m_{\rm j}$ 
to represent the apparent magnitude of the galaxy in this band
and $m_{\rm j,lim}$ to denote the magnitude limit of the survey. 
Following \cite{Pozzetti2010}, we first estimate a mass limit, 
$M_{*,{\rm lim}}$, for each galaxy using 
\begin{equation}
\label{eq:M_lim}
\log(M_{\ast,{\rm lim}})=\log(M_{\ast})+0.4(m_{j}-m_{\rm j,lim})\,
\end{equation}
Galaxies are divided into successive redshift bins with a fixed 
width of $\Delta z=0.2$. For each redshift bin we select the 
20\% faintest galaxies according to the $j$-th band apparent magnitude, 
and obtain the distribution of their $M_{*,{\rm lim}}$.
The minimum stellar mass at a given redshift, $M_{*,{\rm min}}(z)$, is 
then defined by the upper envelope of the distribution, which is the 
value of $M_{*,{\rm lim}}$ that encloses 95\% of the galaxies in 
the distribution. 

The $y$-band GLF and GSMF obtained this way are shown as 
blue triangles in \autoref{fig:pfs_lf_y} and \autoref{fig:pfs_gsmf}, respectively, 
for the PFS-like survey. Panels from left to right are the results for 
three different redshift intervals: $0.7<z<1$, $1<z<1.3$ and $1.3<z<1.6$. 
We take the average of the 20 mock catalogs as 
our mean measurements and their standard deviation as errors. 
The {\em true} GLF and GSMF measured 
from the model galaxies in the simulation are plotted for comparison as 
green symbols and lines. 
The gray vertical dashed line in each panel indicates 
the limiting luminosity ($M_{y,{\rm lim}}$) or the limiting mass 
($M_{\ast,{\rm lim}}$) to which these two functions can be measured
with the PFS-like spectroscopic sample. The limit for the GLF at 
a given redshift is determined by the apparent magnitude limit,
which is either $y<22.5$ or $J<22.8$. For the GSMF, 
we calculate two $M_{\ast, {\rm lim}}$ for each galaxy using 
\autoref{eq:M_lim} --- one for $y_{\rm lim}=22.5$
and the other for $J_{\rm lim}=22.8$, and we use the lower 
one of the two as the actual limit for the galaxy.

As can be seen, above the limiting luminosities and the limiting 
masses, the measurements obtained using the $V_{\rm max}$-weighting 
method can well reproduce the input GLF and GSMF. 
Below the limiting luminosities/masses, however, the measurements 
drop dramatically below the input model owing to the incompleteness 
caused by the flux limit. In order to extend the range 
of the GLF and GSMF measurements, we use the parent photometric 
sample to measure the GLF and GSMF directly, making use of the 
photometric redshift, stellar mass and luminosity of the galaxies in 
the photometric sample which can be estimated from multi-band 
SED fitting. The GLF and GSMF measured from the photometric sample 
are shown as the orange symbols and lines in \autoref{fig:pfs_lf_y} 
and \autoref{fig:pfs_gsmf}, which very well reproduce the model GLF and 
GSMF down to the lowest luminosities/masses and up to the characteristic 
luminosities/masses. At the high-luminosity and high-mass ends, 
both GLF and GSMF are significantly overestimated in the photometric 
sample owing to errors in the luminosity and stellar mass, an effect 
known as the Eddington bias. Our results suggest that one can obtain 
unbiased measurements over the full luminosity or mass range by 
combining the GLF and GSMF measured from the $V_{\rm max}$-weighting 
method above the limiting luminosities/masses and those measured 
from the photometric sample below the limiting luminosities/masses. 

\autoref{fig:pfs_cv_smf} shows the fractional variance in the 
GSMF as estimated from the 20 mock catalogs for the PFS-like survey, 
as a function of the survey area and for different stellar mass 
and redshift intervals. The fractional variance is defined as the 
standard deviation of the GSMF among 20 mocks divided by the mean GSMF of the 20 mocks. 
The PFS-like survey contains 13 pointings. We use a number of pointings in each 
mock survey to calculate the corresponding GSMF, and use the variance among the 20 mocks to obtain 
the corresponding standard deviation. Samples using different number of pointings 
have different survey areas, and we obtain the fractional variance as the function of 
the survey area by using increasing number of pointings from 1 to 13.
The solid lines and dashed lines 
represent results obtained from the $V_{\rm max}$-weighting method 
and the photometric sample, respectively. For comparison, the results 
for the zCOSMOS-bright sample and the VIPERS sample, estimated from 
the corresponding mock catalogs, are plotted as the big hollow and big 
solid symbols, respectively.
Overall, as expected, the total fraction variance in the GSMF 
decreases rapidly with increasing survey area due to the decreasing 
of cosmic variance. 
When compared to the zCOSMOS, the measurement errors are expected to 
decrease from 10\%-20\% to 3\%-6\% for the complete PFS-like survey. 
At fixed survey area, the errors show weak dependence on mass for the two 
lower mass bins, but increase dramatically for the highest mass bin. When the 
survey area exceeds about 5 deg$^2$, the errors continue to decrease, but only
slowly, as the survey size increases. This indicates that the cosmic 
variance is no longer  a dominating source of error when the survey 
area is substantially large. 

\subsection{Projected two-point correlation functions}
\label{sec:cluster}

\begin{figure*}
	\includegraphics[width=\textwidth]{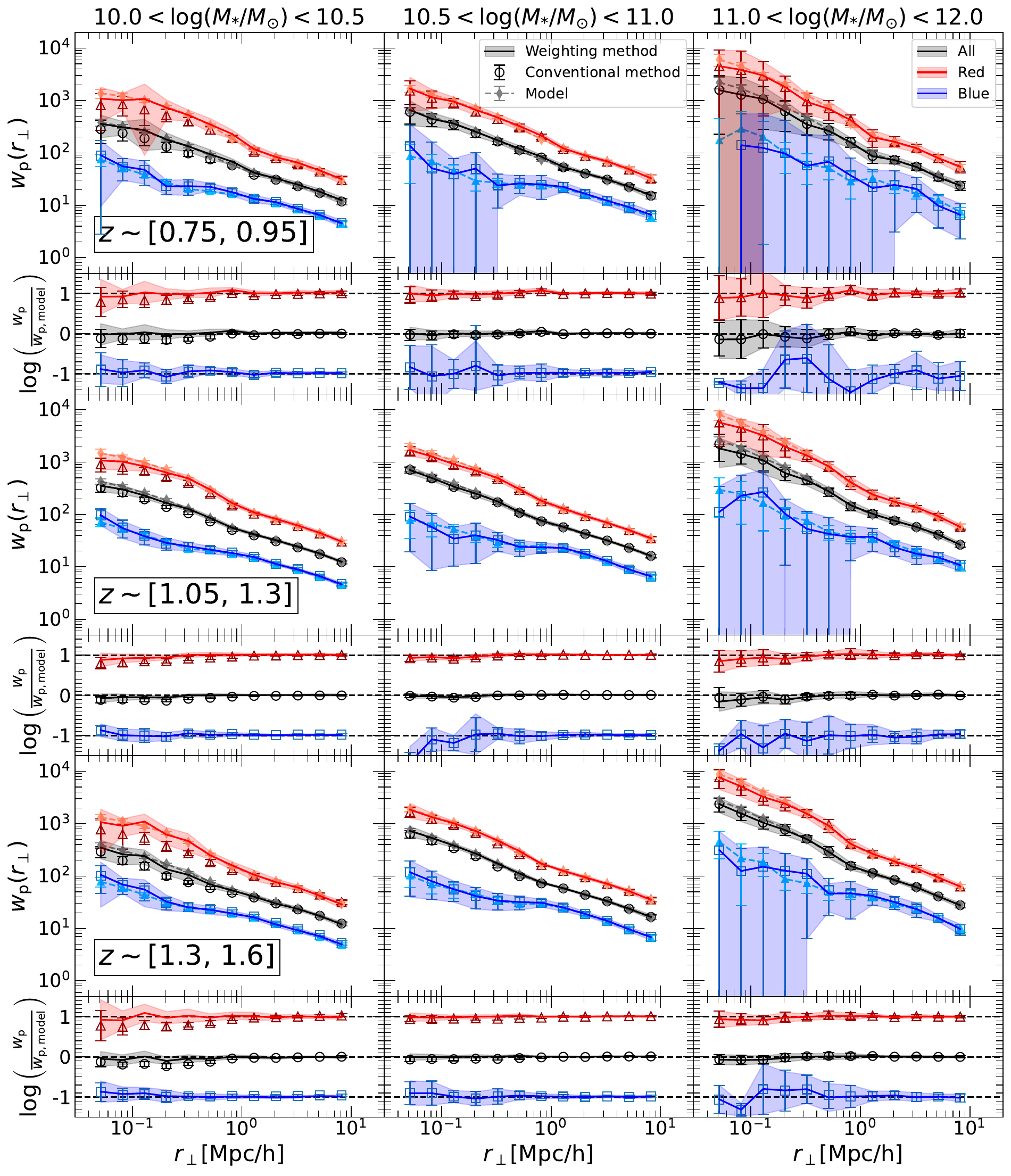}
    \caption{The projected 2PCF $w_{\rm p}(r_{\perp})$ measured from 
    the mock catalogs of the PFS-like galaxy survey, for different 
    redshift intervals and different stellar mass ranges, 
    as indicated. The solid lines represent the $w_{\rm p}(r_{\perp})$ measured with our ``weighting method''. The dashed lines represent the true $w_{\rm p}(r_{\perp})$ for model galaxies. The hollow symbols represent the $w_{\rm p}(r_{\perp})$ measured with ``conventional method''. The black and gray lines/symbols represent the results of the total sample. Red and blue dots and lines represent the results for the subset of red and blue sample. To show the results clearly, we multiply the mean value of red sample and blue sample by 2 and 0.5, and the ratios of red sample and blue sample by 10 and 0.1, respectively.}
    \label{fig:pfs_wp_cor}
\end{figure*}

\begin{figure*}
	\includegraphics[scale=0.5]{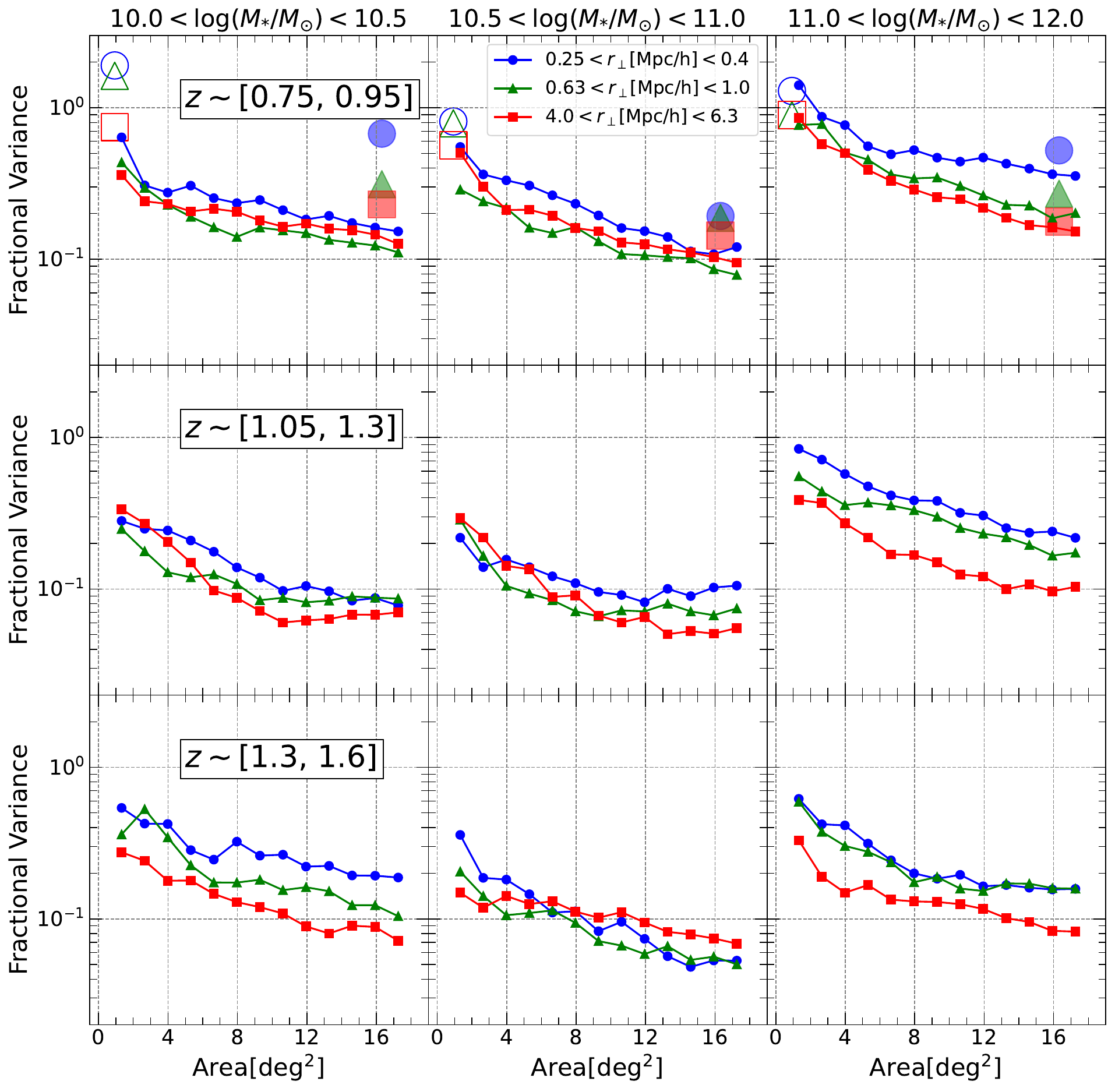}
    \caption{The fractional variance in the projected 
    2PCF is estimated from the mock catalogs of the PFS-like galaxy survey, 
    as a function of the survey area. 
    Results are show for different redshift intervals and different stellar mass intervals in 
    different panels, and in each panel the different symbols/colors 
    are for different $r_{\perp}$ as indicated. The large hollow symbols are the fractional variance estimated 
    for the zCOSMOS-bright survey, and the large solid symbols are the results for VIPERS survey.}
    \label{fig:pfs_cv_wp}
\end{figure*}

We estimate the projected 2PCF $w_{\rm p}(r_{\perp})$ for the mock catalogs 
of the three surveys using the new method proposed in 
\S~\ref{sec:new_method}. Results for the PFS-like survey 
are shown in \autoref{fig:pfs_wp_cor} as solid lines and colored bands which represent errors, 
for three stellar mass intervals and three 
redshift intervals, as indicated in each panel. The black, red and blue solid lines and bands 
represent results for the total sample, and the red and blue samples, respectively. 
Again, the mean measurement of $w_{\rm p}(r_{\perp})$ 
is given by the average of the 20 mock samples, while errors are 
estimated by the standard deviation of the mock samples around the average. 
For the PFS-like survey, the sampling rate $f_{\rm flux}\left(\vec{\mathbf{q}}\right)$ 
caused by the {\em flux limit effect} is not continuous at $z=1$ due to the 
different sample selection criteria below and above this redshift. 
Therefore we do not consider the narrow redshift range of $0.95<z<1.05$
to avoid this problem. For comparison, the {\em true} $w_{\rm p}(r_{\perp})$ 
for the same mass and redshift bins are obtained from model galaxies 
in the simulation and are plotted as the diamonds and solid lines. The projected 2PCF $w_{\rm p}(r_{\perp})$ measured with the  ``conventional method'' are shown as the hollow symbols with error bars. We plot the ratio between 
our measured $w_{\rm p}(r_{\perp})$ with our ``weighting method'' and ``conventional method'' and {\em true} $w_{\rm p}(r_{\perp})$ in 
the lower panels. We shift the ratios for red and blue samples by 1 dex to show the results clearly. The ratios with our ``weighting method'' are around unity in all cases, demonstrating that 
our weighting method can reproduce the input model, for all the mass 
bins and redshift ranges, as well as for both red and blue subsamples. On the other hand, 
the ratios with the ``conventional method'' for total and red samples in the lowest stellar mass bin are 
underestimated at $r_{\perp}<1\ {\rm Mpc/h}$ by 0.1-0.2 dex. Our tests for the zCOSMOS and VIPERS surveys 
are shown in \autoref{sec:other_survey}.

We use the 20 mock catalogs to estimate errors of 
$w_{\rm p}(r_{\perp})$ as functions of the survey area. 
In \autoref{fig:pfs_cv_wp}, we plot the total fractional 
variance among the 20 mock samples, $\sigma_{\rm tot}$, as the 
solid lines, as a function of the survey area for PFS-like survey in three stellar 
mass intervals and three redshift intervals, as indicated in the figure. 
We also show results for the zCOSMOS-bright sample and the 
VIPERS sample as the big hollow and big solid symbols, respectively. 
Different colors represent results for different scales. 
As expected, $\sigma_{\rm tot}$ decreases as 
the survey area increases. However, the decrement is not obvious 
when the survey area exceeds about 8 deg$^2$, similar to the situation 
of the GSMF. This indicates again that cosmic variance is no longer the main 
source of errors when the survey area is large enough, and thus it is not an 
efficient way to reduce the error by further increasing the survey area. 
At large survey area, the errors in the 2PCF measurements should be 
more dominated by short noise caused by limited sampling rates. 
The sampling rate for PFS-like sample is 50\% at $z<1$ and 70\% at $z>1$, 
while the VIPERS survey has a comparable survey area, but has an average 
sampling rate of 30\% only. As can be seen from the figure, the errors 
at $z>1$ is systematically smaller than the errors at $z<1$ at given survey 
area, especially for the two massive stellar mass bins, and the errors for 
the VIPERS sample are also larger than those for the PFS-like sample 
of the same survey area, especially at small scales. 



In summary, the errors in the projected 2PCF for the whole PFS-like sample 
is about 5-20\%, which are a factor of $5\sim 10$ smaller than that of the 
zCOSMOS survey and a factor of $\sim$3 smaller than that of the VIPERS.

\section{Summary}
\label{sec:summary}

In this paper we studied the abundance and clustering 
of galaxies at high redshift as function of luminosity, stellar mass 
and redshift, using realistic mock catalogs constructed 
from cosmological simulations. We populated the 
halos and subhalos of dark matter in the simulation with 
model galaxies of different masses and colors. We calibrate 
the galaxy model using multi-band deep imaging data 
from COSMOS2020. We then constructed mock catalogs 
for spectroscopic surveys based on the model galaxies 
in the simulation. We considered three multi-object 
spectroscopic surveys: the existing zCOSMOS and VIPERS, and  
the PFS-like survey in the future. 
When constructing mock catalogs, we carefully 
included the same selection effects as in the real surveys.
These include area/redshift boundaries, the incomplete 
sampling due to the limited number of fibers/slits and 
the fiber/slit collision effects, and sample selection by 
flux criteria. Using the mock catalogs we measured the 
luminosity function (GLF), stellar mass function (GSMF), 
and projected two-point correlation function (2PCF) for 
galaxies of different stellar mass at different redshift, and we 
quantify the measurement errors using 20 mock samples. 

We found that incomplete sampling can lead to significantly 
biased measurements of galaxy stellar mass function (GSMF) 
and galaxy luminosity function (GLF). We applied a weighting 
scheme to explicitly account for the sky position-dependent sampling rate, 
and our test with the mock catalogs showed that the bias caused 
by the incomplete sampling can be well corrected for the 
GLF and GSMF measurements at masses above the characteristic 
luminosity/mass. In addition, we use the parent photometric sample 
to extend the range of the measured GLF and GSMF down to 
luminosities/masses well below the limits of the spectroscopic sample. 

For galaxy clustering, we found that both the 
target selection by flux criteria and the incomplete 
sampling can lead to significantly biased measurements. 
The effect of incomplete sampling 
can be fully corrected using our weighting scheme and the correction 
of fiber/slit collision (\autoref{fig:wp_cor}), but the 
bias due to the selection criteria of flux-limited samples 
cannot be corrected in a simple way. 
The bias caused by the flux limit effect is more
significant at lower masses and at higher redshifts, and 
is mainly contributed by the population of low-mass red 
galaxies which are more strongly clustered but less well 
sampled due to their larger mass-to-light ratios in comparison 
to blue galaxies. We developed a weighting method to correct this bias, 
with the help of two photometric samples that can be obtained 
from the parent data to estimate the sampling rate caused by the 
flux limit. We showed that the bias in the projected $w_{\rm p}(r_{\perp})$ 
can be corrected with this method, for all the stellar mass and 
redshift ranges considered, as well as separately for 
red and blue samples (\autoref{fig:pfs_wp_cor}). 

We used the mock catalogs to estimate errors in 
both the abundance and clustering measurements,
and we made comparisons between zCOSMOS, VIPERS and the PFS-like 
galaxy survey. The fractional variance of the GSMF decreases with survey area, 
from 10-20\% in zCOSMOS-like surveys that have a survey 
area of $\sim $ 1 deg$^2$ down to 3-6\% in PFS-like surveys 
that covers $\sim14.5$ ${\rm deg}^2$ (\autoref{fig:pfs_cv_smf}). 
For the projected 2PCF, the fractional variance also decreases with survey area, 
from $>50\%$ in zCOSMOS-like surveys, 10\%-70\% in VIPERS-like surveys 
to 5-20\% in PFS-like surveys. We find that, when the survey area is 
sufficiently large ($\sim$4 deg$^2$ for GSMF and $\sim$8 deg$^2$ for 2PCF), 
the errors decrease only slowly as survey area increases. This result 
indicates that the cosmic variance is no longer the dominant source of error 
when the survey area exceeds those numbers. At such large survey area,
our results indicate that the more efficient way to reduce errors is
to increase sampling rate rather than survey area.

Our analyses are based on three surveys with quite different sample selections. 
The fact that our methods work accurately for all of them indicate that 
the main test results are valid for other redshift surveys. In particular, 
by using information from the parent photometric survey, one can obtain 
unbiased measurements for both the luminosity/stellar mass function and 
the correlation function for different populations of galaxies. With the 
advent of deep redshift surveys of galaxies, our results provide an 
important guideline to analyze the observational data. 

All the mock catalogs for zCOSMOS, VIPERS and PFS-like surveys 
constructed in this work are publicly available in \citet{Mock_data} and \href{https://lig.astro.tsinghua.edu.cn/astrodata}
{https://lig.astro.tsinghua.edu.cn/astrodata}. 

\begin{acknowledgments}
  This work is supported by the National Key R\&D Program of China
  (grant No. 2018YFA0404502), and the National Science 
  Foundation of China (grant Nos. 11821303, 11733002, 11973030, 
  11673015, 11733004, 11761131004, 11761141012).
Part of our analysis is based on data products from observations made with
ESO Telescopes at the La Silla Paranal Observatory under ESO
programme ID 179.A-2005 and on data products produced by TERAPIX
and the Cambridge Astronomy Survey Unit on behalf of the
UltraVISTA consortium.

The authors acknowledge the Tsinghua Astrophysics High-Performance Computing
platform at Tsinghua University for providing computational and data storage resources
that have contributed to the research results reported within this paper.

\end{acknowledgments}

%



\software{Astropy \citep{2013A&A...558A..33A,2018AJ....156..123A},  
          NumPy \citep{numpy2020}, 
          SciPy \citep{scipy2020},
          Matplotlib \citep{Matplotlib2007},
          h5py \citep{h5py2021},
          emcee \citep{emcee2013},
          CIGALE \citep{Boquien2019},
          HIPP \citep{HIPP}}


\appendix

\section{Spatial sampling rates versus surface number density of targets}
\label{sec:sample_rate}

\begin{figure}
	\includegraphics[width=\columnwidth]{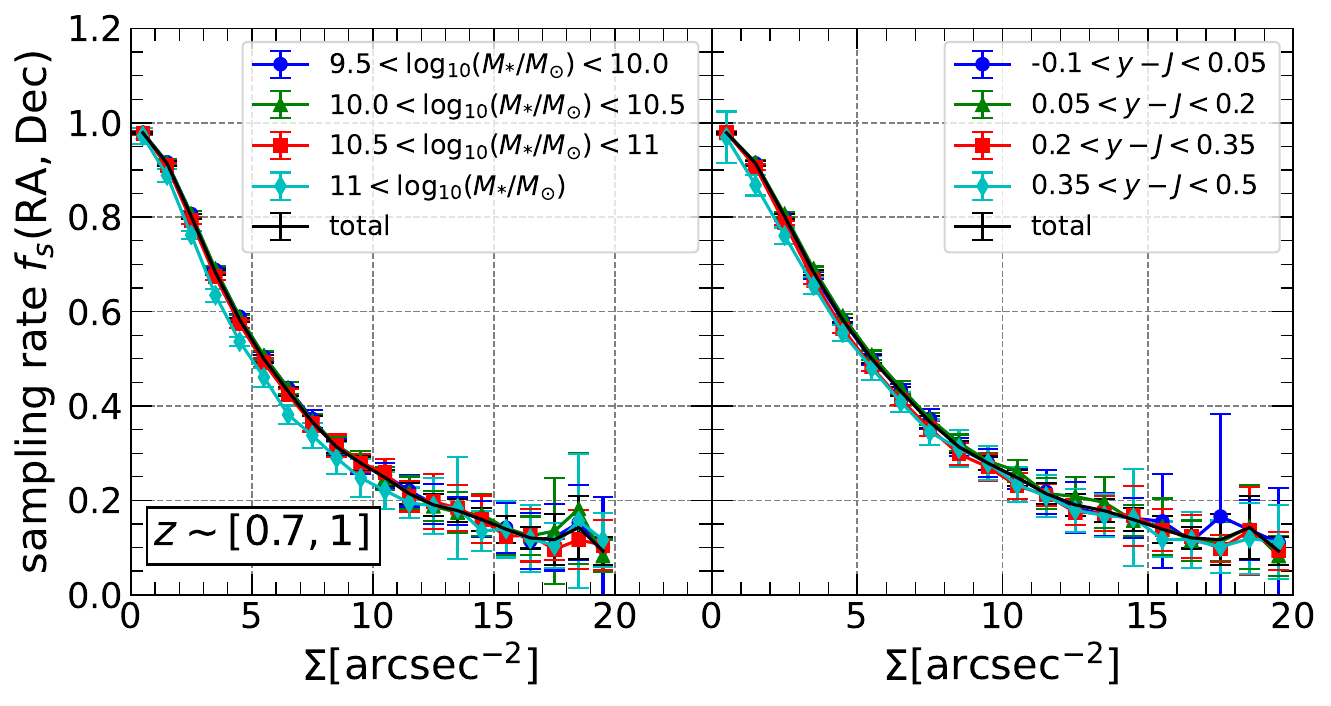}
    \caption{The local sampling rate as a 
    function of the local surface number 
    density of galaxy targets. Color lines represent  
    results for different stellar mass ranges, while the 
    black line is the result for galaxies as a whole.}
    \label{fig:sample_rate}
\end{figure}

As shown in the main text, the spatial incomplete sampling of 
galaxy targets in PFS-like galaxy samples 
can well be corrected using our weighting scheme. 
This is mainly because the spatial sampling rate depends 
only on the local surface number density of the galaxy 
targets, although the sampling may vary from region to region. 
This is demonstrated clearly in \autoref{fig:sample_rate}, 
where we plot the sky position-dependent sampling 
rate, $f_{s}({\rm RA,Dec})$, in the PFS-like mock catalogs, 
defined in \S~\ref{sec:conventional_method}, 
as a function of the surface density $\Sigma$. 
Here, for each galaxy in a mock catalog, we have 
estimated a local surface density $\Sigma$ by counting the 
number of galaxies in the parent photometric sample 
that are located within 1 arcmin. We show the sampling 
rate versus $\Sigma$ for subsamples selected by the stellar 
mass (left panel) or the $y-J$ color (right panel).
All the measurements overlap with each other, and 
are almost the same as the result of the full sample
(the black curve), which is the average of the 20 mock 
catalogs constructed in \S~\ref{sec:mock_construction}. 

The figure shows that the sampling rate decreases rapidly 
with $\Sigma$, varying from nearly 100\% 
in the lowest density regions down to $\sim$10\% in 
the highest. As shown in the main text, this variation 
causes biases in the estimation of the luminosity function, 
stellar mass function and correlation function. 
However, our weighting scheme is able to correct these 
biases for all the statistics even when the sampling 
rate is very low. 

\section{The errors of the project 2PCF as function of survey area and sampling rate}
\label{sec:wp_error_area_sr}

\begin{figure*}
	\includegraphics[width=\textwidth]{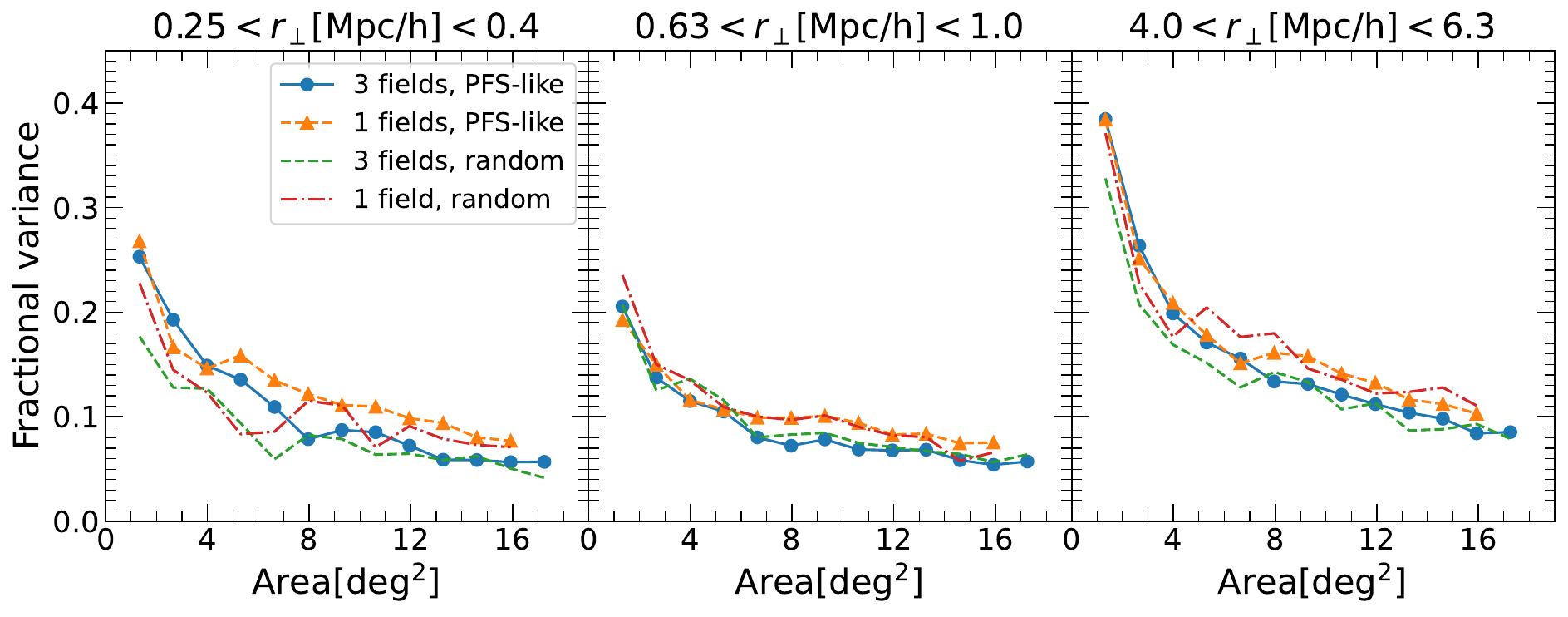}
	\includegraphics[width=\textwidth]{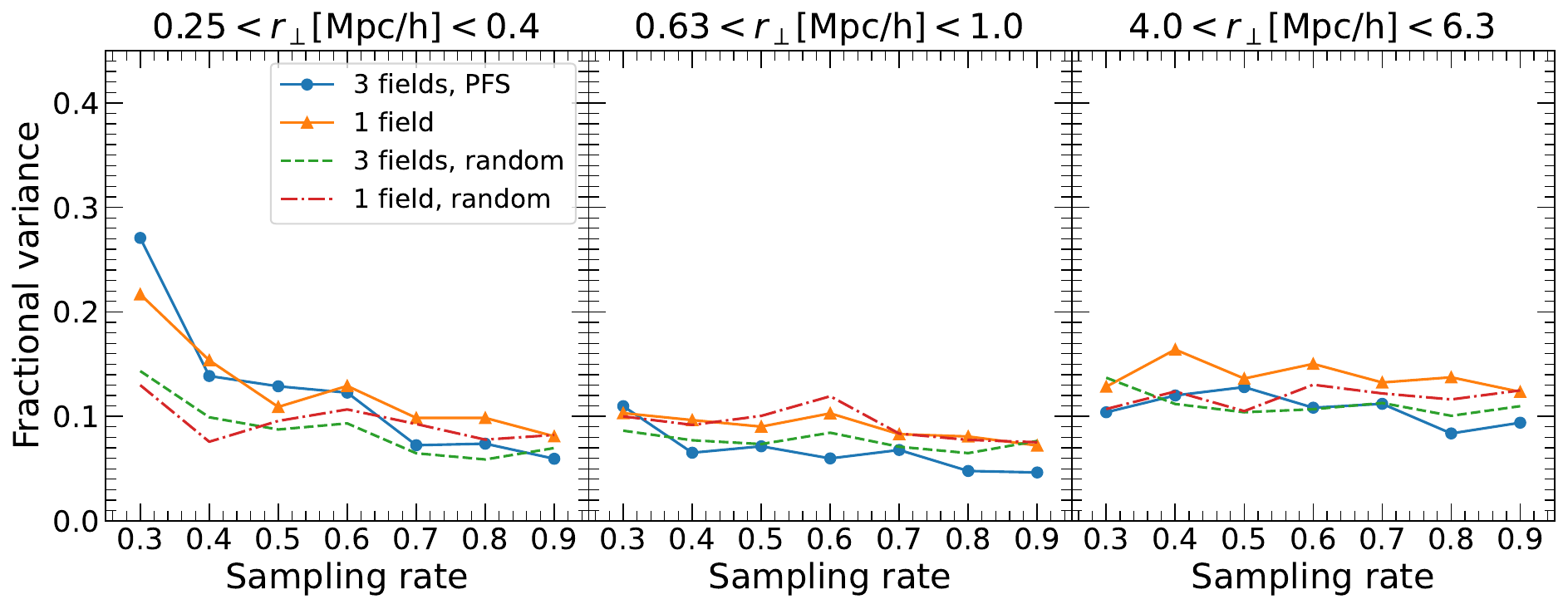}
	\caption{The fractional variances as function of survey area and sampling rate at the redshift range $0.75<z<0.95$ and in stellar mass interval $10.5<\log\left(M_{*}/M_{\odot}\right)<11.0$. The blue dots and orange triangles represent the results for the PFS-like surveys using PFS target selection strategy with 3 separated fields and 1 field. The dashed and dot dashed lines represent the results for the random sampling with 3 separated fields and 1 field.}
	\label{fig:wp_error_area_sr}
\end{figure*}

In the main text we find that errors in the measurements of the 
projected 2PCF depend on both the survey area and the sampling rate, 
but in different ways. Here we present more results of this analysis. 
We consider the redshift range $0.75<z<0.95$ and the stellar mass interval
$10.5<\log\left(M_{*}/M_{\odot}\right)<11.0$ in the PFS-like survey as an example. 
\autoref{fig:wp_error_area_sr} shows the fractional variance of the 
measurements from the 20 mock catalogs as functions of survey area 
(upper panels) and sampling rate (lower panels), 
We have fixed the sampling rate at 70\% for the upper panels 
and the survey area at 12 deg$^2$ for the lower panel, respectively, 
in order to separate effects of the two parameters. Panels 
from left to right show results for three different $r_{\perp}$ scales. 
Plotted in the blue symbols/lines are the results for the PFS-like survey 
considered in the main text, where the total survey area is divided 
into three fields that consist of similar numbers of pointings but 
are well separated in the sky. The yellow symbols/lines show the 
case in which the survey has only a single field, but with the 
same total area and fiber assignment as the PFS-like survey. 
In addition, the green dashed line and the red dashed-dotted line 
in each panel show the cases with three separate fields and one 
single field, respectively, but with purely random sampling 
(i.e. without considering any fiber assignment). The PFS-like survey 
with three separate fields gives smaller errors than the one with 
a single field, particularly when the survey area is above $\sim5$ deg$^2$ 
and the sampling rate is above $\sim0.4$. This is because the cosmic 
variance is expected to be smaller in the case of three separated 
fields. Random sampling results in smaller errors than the PFS-like 
surveys, implying that fiber collisions introduce additional errors, 
particularly at small scales and when the sampling rate is low 
(see the left bottom panel). In a recent study, 
\citet{Pearl2022} constructed mock catalogs for the PFS galaxy 
evolution survey assuming a single field and random sampling. Our 
results are consistent with theirs in this case. However, since 
real surveys such as the PFS always use fibers (or slitless masks), 
it is necessary to include effects of fiber assignment in the mock 
catalog if one were to accurately model bias and error in 
clustering measurements. 

\section{Projected two-point correlation functions in zCOSMOS and VIPERS survey}
\label{sec:other_survey}

\begin{figure*}
	\includegraphics[width=\textwidth]{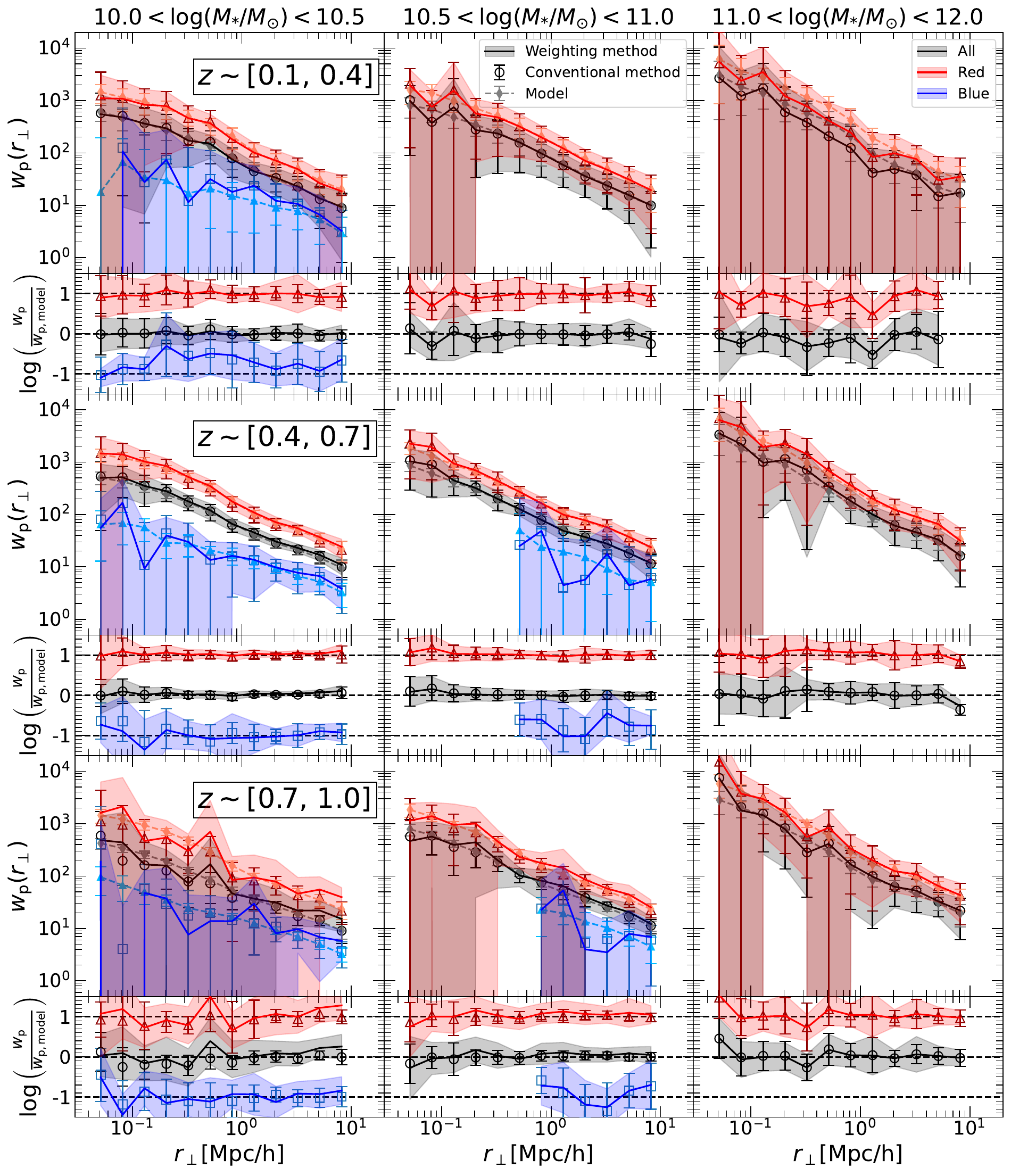}
    \caption{The projected 2PCF $w_{\rm p}(r_{\perp})$ measured from 
    the mock catalogs of the zCOSMOS survey, for different 
    redshift intervals and different stellar mass ranges, 
    as indicated. The solid lines represent the $w_{\rm p}(r_{\perp})$ measured with our ``weighting method''. The dashed lines represent the true $w_{\rm p}(r_{\perp})$ for model galaxies. The hollow symbols represent the $w_{\rm p}(r_{\perp})$ measured with ``conventional method''. The black and gray lines/symbols represent the results of the total sample. Red and blue dots and lines represent the results for the subset of red and blue sample. To show the results clearly, we multiply the mean value of red sample and blue sample by 2 and 0.5, and the ratios of red sample and blue sample by 10 and 0.1, respectively.}
    \label{fig:zc_wp_cor}
\end{figure*}

\begin{figure*}
	\includegraphics[width=\textwidth]{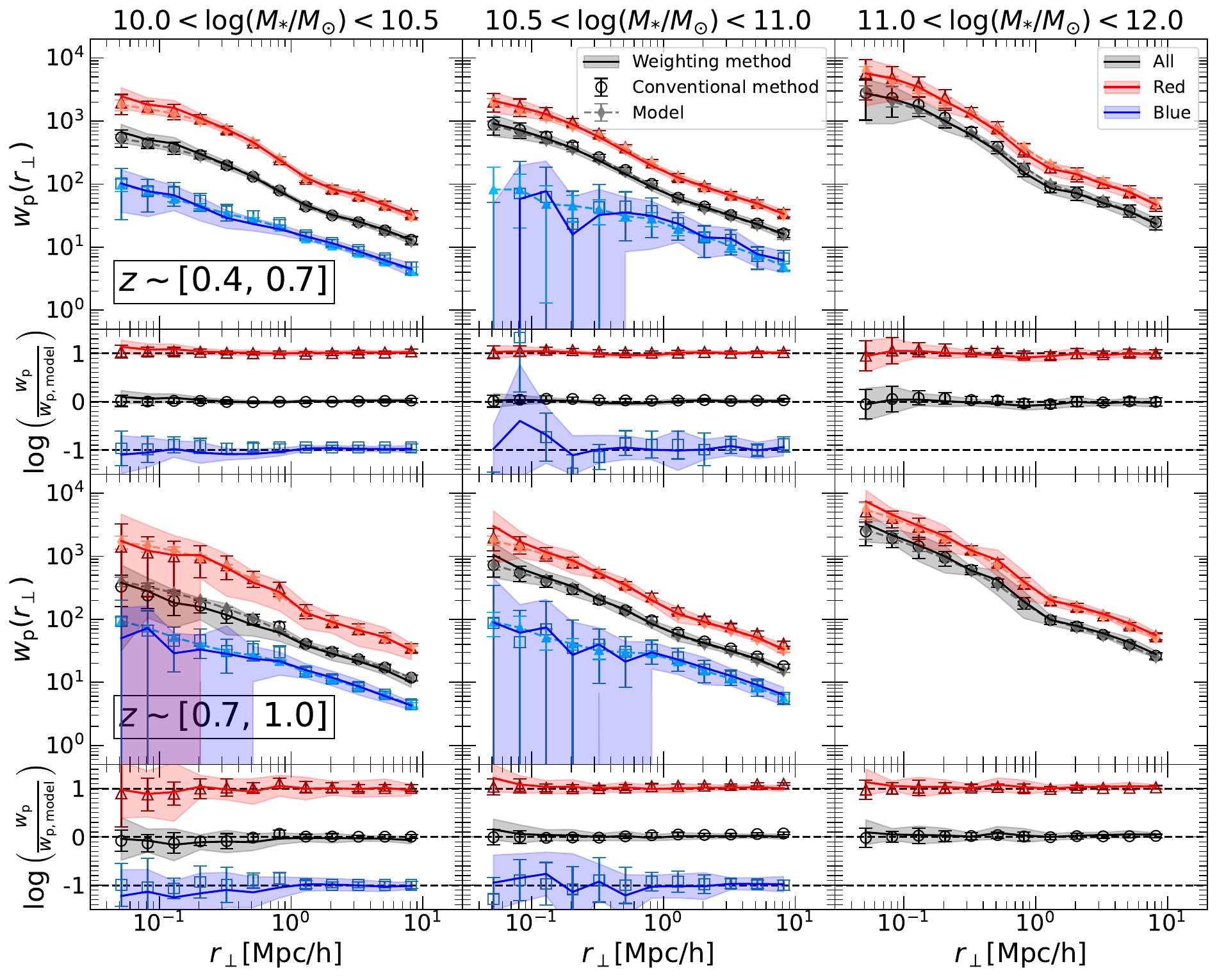}
    \caption{The projected 2PCF $w_{\rm p}(r_{\perp})$ measured from 
    the mock catalogs of the VIPERS survey, for different 
    redshift intervals and different stellar mass ranges, 
    as indicated. The solid lines represent the $w_{\rm p}(r_{\perp})$ measured with our ``weighting method''. The dashed lines represent the true $w_{\rm p}(r_{\perp})$ for model galaxies. The hollow symbols represent the $w_{\rm p}(r_{\perp})$ measured with ``conventional method''. The black and gray lines/symbols represent the results of the total sample. Red and blue dots and lines represent the results for the subset of red and blue sample. To show the results clearly, we multiply the mean value of red sample and blue sample by 2 and 0.5, and the ratios of red sample and blue sample by 10 and 0.1, respectively.}
    \label{fig:vipers_wp_cor}
\end{figure*}

We estimate the projected 2PCF $w_{\rm p}(r_{\perp})$ for mock catalogs of the three surveys, 
the PFS-like, zCOSMOS and VIPERS. The results are shown in \autoref{fig:zc_wp_cor} and \autoref{fig:vipers_wp_cor} 
for zCOSMOS and VIPERS survey, respectively. The coding of lines and symbols is the same as that in 
\autoref{fig:pfs_wp_cor} and is described in \autoref{sec:cluster}. We find that the ratios between the 
measured $w_{\rm p}(r_{\perp})$ and model results are all around unity for both the ``weighting method'' 
and the ``conventional method'', indicating that the {\em flux limit effect} is not significant in 
surveys like zCOSMOS and VIPERS. Some panels lack results for blue galaxies,
because the samples are too small.

\bibliography{sample631}{}

\begin{thebibliography}{}
\expandafter\ifx\csname natexlab\endcsname\relax\def\natexlab#1{#1}\fi
\providecommand{\url}[1]{\href{#1}{#1}}
\providecommand{\dodoi}[1]{doi:~\href{http://doi.org/#1}{\nolinkurl{#1}}}
\providecommand{\doeprint}[1]{\href{http://ascl.net/#1}{\nolinkurl{http://ascl.net/#1}}}
\providecommand{\doarXiv}[1]{\href{https://arxiv.org/abs/#1}{\nolinkurl{https://arxiv.org/abs/#1}}}

\bibitem[{{Abbas} \& {Sheth}(2006)}]{Abbas2006}
{Abbas}, U., \& {Sheth}, R.~K. 2006, \mnras, 372, 1749,
  \dodoi{10.1111/j.1365-2966.2006.10987.x}

\bibitem[{{Aihara} {et~al.}(2018){Aihara}, {Arimoto}, {Armstrong}, {Arnouts},
  {Bahcall}, {Bickerton}, {Bosch}, {Bundy}, {Capak}, {Chan}, {Chiba}, {Coupon},
  {Egami}, {Enoki}, {Finet}, {Fujimori}, {Fujimoto}, {Furusawa}, {Furusawa},
  {Goto}, {Goulding}, {Greco}, {Greene}, {Gunn}, {Hamana}, {Harikane},
  {Hashimoto}, {Hattori}, {Hayashi}, {Hayashi}, {He{\l}miniak}, {Higuchi},
  {Hikage}, {Ho}, {Hsieh}, {Huang}, {Huang}, {Ikeda}, {Imanishi}, {Inoue},
  {Iwasawa}, {Iwata}, {Jaelani}, {Jian}, {Kamata}, {Karoji}, {Kashikawa},
  {Katayama}, {Kawanomoto}, {Kayo}, {Koda}, {Koike}, {Kojima}, {Komiyama},
  {Konno}, {Koshida}, {Koyama}, {Kusakabe}, {Leauthaud}, {Lee}, {Lin}, {Lin},
  {Lupton}, {Mand elbaum}, {Matsuoka}, {Medezinski}, {Mineo}, {Miyama},
  {Miyatake}, {Miyazaki}, {Momose}, {More}, {More}, {Moritani}, {Moriya},
  {Morokuma}, {Mukae}, {Murata}, {Murayama}, {Nagao}, {Nakata}, {Niida},
  {Niikura}, {Nishizawa}, {Obuchi}, {Oguri}, {Oishi}, {Okabe}, {Okamoto},
  {Okura}, {Ono}, {Onodera}, {Onoue}, {Osato}, {Ouchi}, {Price}, {Pyo}, {Sako},
  {Sawicki}, {Shibuya}, {Shimasaku}, {Shimono}, {Shirasaki}, {Silverman},
  {Simet}, {Speagle}, {Spergel}, {Strauss}, {Sugahara}, {Sugiyama}, {Suto},
  {Suyu}, {Suzuki}, {Tait}, {Takada}, {Takata}, {Tamura}, {Tanaka}, {Tanaka},
  {Tanaka}, {Tanaka}, {Terai}, {Terashima}, {Toba}, {Tominaga}, {Toshikawa},
  {Turner}, {Uchida}, {Uchiyama}, {Umetsu}, {Uraguchi}, {Urata}, {Usuda},
  {Utsumi}, {Wang}, {Wang}, {Wong}, {Yabe}, {Yamada}, {Yamanoi}, {Yasuda},
  {Yeh}, {Yonehara}, \& {Yuma}}]{Aihara2018}
{Aihara}, H., {Arimoto}, N., {Armstrong}, R., {et~al.} 2018, \pasj, 70, S4,
  \dodoi{10.1093/pasj/psx066}

\bibitem[{{Artale} {et~al.}(2017){Artale}, {Pedrosa}, {Trayford}, {Theuns},
  {Farrow}, {Norberg}, {Zehavi}, {Bower}, \& {Schaller}}]{Artale_etal2017}
{Artale}, M.~C., {Pedrosa}, S.~E., {Trayford}, J.~W., {et~al.} 2017, \mnras,
  470, 1771, \dodoi{10.1093/mnras/stx1263}

\bibitem[{{Astropy Collaboration} {et~al.}(2013){Astropy Collaboration},
  {Robitaille}, {Tollerud}, {Greenfield}, {Droettboom}, {Bray}, {Aldcroft},
  {Davis}, {Ginsburg}, {Price-Whelan}, {Kerzendorf}, {Conley}, {Crighton},
  {Barbary}, {Muna}, {Ferguson}, {Grollier}, {Parikh}, {Nair}, {Unther},
  {Deil}, {Woillez}, {Conseil}, {Kramer}, {Turner}, {Singer}, {Fox}, {Weaver},
  {Zabalza}, {Edwards}, {Azalee Bostroem}, {Burke}, {Casey}, {Crawford},
  {Dencheva}, {Ely}, {Jenness}, {Labrie}, {Lim}, {Pierfederici}, {Pontzen},
  {Ptak}, {Refsdal}, {Servillat}, \& {Streicher}}]{2013A&A...558A..33A}
{Astropy Collaboration}, {Robitaille}, T.~P., {Tollerud}, E.~J., {et~al.} 2013,
  \aap, 558, A33, \dodoi{10.1051/0004-6361/201322068}

\bibitem[{{Astropy Collaboration} {et~al.}(2018){Astropy Collaboration},
  {Price-Whelan}, {Sip{\H{o}}cz}, {G{\"u}nther}, {Lim}, {Crawford}, {Conseil},
  {Shupe}, {Craig}, {Dencheva}, {Ginsburg}, {VanderPlas}, {Bradley},
  {P{\'e}rez-Su{\'a}rez}, {de Val-Borro}, {Aldcroft}, {Cruz}, {Robitaille},
  {Tollerud}, {Ardelean}, {Babej}, {Bach}, {Bachetti}, {Bakanov}, {Bamford},
  {Barentsen}, {Barmby}, {Baumbach}, {Berry}, {Biscani}, {Boquien}, {Bostroem},
  {Bouma}, {Brammer}, {Bray}, {Breytenbach}, {Buddelmeijer}, {Burke},
  {Calderone}, {Cano Rodr{\'\i}guez}, {Cara}, {Cardoso}, {Cheedella}, {Copin},
  {Corrales}, {Crichton}, {D'Avella}, {Deil}, {Depagne}, {Dietrich}, {Donath},
  {Droettboom}, {Earl}, {Erben}, {Fabbro}, {Ferreira}, {Finethy}, {Fox},
  {Garrison}, {Gibbons}, {Goldstein}, {Gommers}, {Greco}, {Greenfield},
  {Groener}, {Grollier}, {Hagen}, {Hirst}, {Homeier}, {Horton}, {Hosseinzadeh},
  {Hu}, {Hunkeler}, {Ivezi{\'c}}, {Jain}, {Jenness}, {Kanarek}, {Kendrew},
  {Kern}, {Kerzendorf}, {Khvalko}, {King}, {Kirkby}, {Kulkarni}, {Kumar},
  {Lee}, {Lenz}, {Littlefair}, {Ma}, {Macleod}, {Mastropietro}, {McCully},
  {Montagnac}, {Morris}, {Mueller}, {Mumford}, {Muna}, {Murphy}, {Nelson},
  {Nguyen}, {Ninan}, {N{\"o}the}, {Ogaz}, {Oh}, {Parejko}, {Parley}, {Pascual},
  {Patil}, {Patil}, {Plunkett}, {Prochaska}, {Rastogi}, {Reddy Janga},
  {Sabater}, {Sakurikar}, {Seifert}, {Sherbert}, {Sherwood-Taylor}, {Shih},
  {Sick}, {Silbiger}, {Singanamalla}, {Singer}, {Sladen}, {Sooley},
  {Sornarajah}, {Streicher}, {Teuben}, {Thomas}, {Tremblay}, {Turner},
  {Terr{\'o}n}, {van Kerkwijk}, {de la Vega}, {Watkins}, {Weaver}, {Whitmore},
  {Woillez}, {Zabalza}, \& {Astropy Contributors}}]{2018AJ....156..123A}
{Astropy Collaboration}, {Price-Whelan}, A.~M., {Sip{\H{o}}cz}, B.~M., {et~al.}
  2018, \aj, 156, 123, \dodoi{10.3847/1538-3881/aabc4f}

\bibitem[{{Baldry} {et~al.}(2008){Baldry}, {Glazebrook}, \&
  {Driver}}]{Baldry2008}
{Baldry}, I.~K., {Glazebrook}, K., \& {Driver}, S.~P. 2008, \mnras, 388, 945,
  \dodoi{10.1111/j.1365-2966.2008.13348.x}

\bibitem[{{Baldry} {et~al.}(2012){Baldry}, {Driver}, {Loveday}, {Taylor},
  {Kelvin}, {Liske}, {Norberg}, {Robotham}, {Brough}, {Hopkins}, {Bamford},
  {Peacock}, {Bland-Hawthorn}, {Conselice}, {Croom}, {Jones}, {Parkinson},
  {Popescu}, {Prescott}, {Sharp}, \& {Tuffs}}]{Baldry2012}
{Baldry}, I.~K., {Driver}, S.~P., {Loveday}, J., {et~al.} 2012, \mnras, 421,
  621, \dodoi{10.1111/j.1365-2966.2012.20340.x}

\bibitem[{{Behroozi} {et~al.}(2013){Behroozi}, {Wechsler}, {Wu}, {Busha},
  {Klypin}, \& {Primack}}]{Behroozi2013}
{Behroozi}, P.~S., {Wechsler}, R.~H., {Wu}, H.-Y., {et~al.} 2013, \apj, 763,
  18, \dodoi{10.1088/0004-637X/763/1/18}

\bibitem[{{Bell} {et~al.}(2003){Bell}, {McIntosh}, {Katz}, \&
  {Weinberg}}]{Bell2003}
{Bell}, E.~F., {McIntosh}, D.~H., {Katz}, N., \& {Weinberg}, M.~D. 2003, \apjs,
  149, 289, \dodoi{10.1086/378847}

\bibitem[{{Bianchi} \& {Percival}(2017)}]{Bianchi2017}
{Bianchi}, D., \& {Percival}, W.~J. 2017, \mnras, 472, 1106,
  \dodoi{10.1093/mnras/stx2053}

\bibitem[{{Bianchi} {et~al.}(2018){Bianchi}, {Burden}, {Percival}, {Brooks},
  {Cahn}, {Forero-Romero}, {Levi}, {Ross}, \& {Tarle}}]{Bianchi2018}
{Bianchi}, D., {Burden}, A., {Percival}, W.~J., {et~al.} 2018, \mnras, 481,
  2338, \dodoi{10.1093/mnras/sty2377}

\bibitem[{{Blaizot} {et~al.}(2005){Blaizot}, {Wadadekar}, {Guiderdoni},
  {Colombi}, {Bertin}, {Bouchet}, {Devriendt}, \& {Hatton}}]{Blaizot2005}
{Blaizot}, J., {Wadadekar}, Y., {Guiderdoni}, B., {et~al.} 2005, \mnras, 360,
  159, \dodoi{10.1111/j.1365-2966.2005.09019.x}

\bibitem[{{Boquien} {et~al.}(2019){Boquien}, {Burgarella}, {Roehlly}, {Buat},
  {Ciesla}, {Corre}, {Inoue}, \& {Salas}}]{Boquien2019}
{Boquien}, M., {Burgarella}, D., {Roehlly}, Y., {et~al.} 2019, \aap, 622, A103,
  \dodoi{10.1051/0004-6361/201834156}

\bibitem[{{Bundy} {et~al.}(2003){Bundy}, {Conselice}, {Ellis}, {Eisenhardt}, \&
  {DEEP2 Team}}]{Bundy2003}
{Bundy}, K., {Conselice}, C., {Ellis}, R., {Eisenhardt}, P., \& {DEEP2 Team}.
  2003, in American Astronomical Society Meeting Abstracts, Vol. 203, American
  Astronomical Society Meeting Abstracts, 106.06

\bibitem[{{Chen} {et~al.}(2019){Chen}, {Mo}, {Li}, {Wang}, {Yang}, {Zhou}, \&
  {Zhang}}]{Yangyao2019}
{Chen}, Y., {Mo}, H.~J., {Li}, C., {et~al.} 2019, \apj, 872, 180,
  \dodoi{10.3847/1538-4357/ab0208}

\bibitem[{{Chen} \& {Wang}(2023)}]{HIPP}
{Chen}, Y., \& {Wang}, K. 2023, {HIPP: HIgh-Performance Package for scientific
  computation}, Astrophysics Source Code Library, record ascl:2301.030.
\newblock \doeprint{2301.030}

\bibitem[{{Coil} {et~al.}(2006){Coil}, {Newman}, {Cooper}, {Davis}, {Faber},
  {Koo}, \& {Willmer}}]{Coil2006}
{Coil}, A.~L., {Newman}, J.~A., {Cooper}, M.~C., {et~al.} 2006, \apj, 644, 671,
  \dodoi{10.1086/503601}

\bibitem[{{Cole} {et~al.}(2000){Cole}, {Lacey}, {Baugh}, \& {Frenk}}]{Cole2000}
{Cole}, S., {Lacey}, C.~G., {Baugh}, C.~M., \& {Frenk}, C.~S. 2000, \mnras,
  319, 168, \dodoi{10.1046/j.1365-8711.2000.03879.x}

\bibitem[{{Cole} {et~al.}(2001){Cole}, {Norberg}, {Baugh}, {Frenk},
  {Bland-Hawthorn}, {Bridges}, {Cannon}, {Colless}, {Collins}, {Couch},
  {Cross}, {Dalton}, {De Propris}, {Driver}, {Efstathiou}, {Ellis},
  {Glazebrook}, {Jackson}, {Lahav}, {Lewis}, {Lumsden}, {Maddox}, {Madgwick},
  {Peacock}, {Peterson}, {Sutherland}, \& {Taylor}}]{Cole2001}
{Cole}, S., {Norberg}, P., {Baugh}, C.~M., {et~al.} 2001, \mnras, 326, 255,
  \dodoi{10.1046/j.1365-8711.2001.04591.x}

\bibitem[{{Collette} {et~al.}(2021){Collette}, {Kluyver}, {Caswell},
  {Tocknell}, {Kieffer}, {Jelenak}, {Scopatz}, {Dale}, {Chen}, {VINCENT},
  {Payno}, {Juliagarriga}, {Sciarelli}, {Valls}, {Ghosh}, {Kofoed Pedersen},
  {Jakirkham}, {Raspaud}, {Danilevski}, {Abbasi}, {Readey}, {Paramonov},
  {Chan}, {Sol{\'e}}, {Jialin}, {Feng}, {Vaillant}, {Teichmann}, {Brucher}, \&
  {Johnson}}]{h5py2021}
{Collette}, A., {Kluyver}, T., {Caswell}, T.~A., {et~al.} 2021, {h5py/h5py:
  3.5.0}, 3.5.0, Zenodo,  Zenodo, \dodoi{10.5281/zenodo.5585380}

\bibitem[{{Davidzon} {et~al.}(2013){Davidzon}, {Bolzonella}, {Coupon},
  {Ilbert}, {Arnouts}, {de la Torre}, {Fritz}, {De Lucia}, {Iovino}, {Granett},
  {Zamorani}, {Guzzo}, {Abbas}, {Adami}, {Bel}, {Bottini}, {Branchini},
  {Cappi}, {Cucciati}, {Franzetti}, {Fumana}, {Garilli}, {Krywult}, {Le Brun},
  {Le F{\`e}vre}, {Maccagni}, {Ma{\l}ek}, {Marulli}, {McCracken}, {Paioro},
  {Peacock}, {Polletta}, {Pollo}, {Schlagenhaufer}, {Scodeggio}, {Tasca},
  {Tojeiro}, {Vergani}, {Zanichelli}, {Burden}, {Di Porto}, {Marchetti},
  {Marinoni}, {Mellier}, {Moscardini}, {Moutard}, {Nichol}, {Percival},
  {Phleps}, \& {Wolk}}]{Davidzon2013}
{Davidzon}, I., {Bolzonella}, M., {Coupon}, J., {et~al.} 2013, \aap, 558, A23,
  \dodoi{10.1051/0004-6361/201321511}

\bibitem[{{Davis} \& {Peebles}(1983)}]{Davis1983}
{Davis}, M., \& {Peebles}, P.~J.~E. 1983, \apj, 267, 465,
  \dodoi{10.1086/160884}

\bibitem[{{de la Torre} {et~al.}(2011){de la Torre}, {Le F{\`e}vre},
  {Porciani}, {Guzzo}, {Meneux}, {Abbas}, {Tasca}, {Carollo}, {Contini}, \&
  {Kneib}}]{Torre2011}
{de la Torre}, S., {Le F{\`e}vre}, O., {Porciani}, C., {et~al.} 2011, \mnras,
  412, 825, \dodoi{10.1111/j.1365-2966.2010.17939.x}

\bibitem[{{Diener} {et~al.}(2013){Diener}, {Lilly}, {Knobel}, {Zamorani},
  {Lemson}, {Kampczyk}, {Scoville}, {Carollo}, {Contini}, {Kneib}, {Le Fevre},
  {Mainieri}, {Renzini}, {Scodeggio}, {Bardelli}, {Bolzonella}, {Bongiorno},
  {Caputi}, {Cucciati}, {de la Torre}, {de Ravel}, {Franzetti}, {Garilli},
  {Iovino}, {Kova{\v{c}}}, {Lamareille}, {Le Borgne}, {Le Brun}, {Maier},
  {Mignoli}, {Pello}, {Peng}, {Perez Montero}, {Presotto}, {Silverman},
  {Tanaka}, {Tasca}, {Tresse}, {Vergani}, {Zucca}, {Bordoloi}, {Cappi},
  {Cimatti}, {Coppa}, {Koekemoer}, {L{\'o}pez-Sanjuan}, {McCracken}, {Moresco},
  {Nair}, {Pozzetti}, \& {Welikala}}]{Diener2013}
{Diener}, C., {Lilly}, S.~J., {Knobel}, C., {et~al.} 2013, \apj, 765, 109,
  \dodoi{10.1088/0004-637X/765/2/109}

\bibitem[{{Driver} \& {Robotham}(2010)}]{Driver2010}
{Driver}, S.~P., \& {Robotham}, A. S.~G. 2010, \mnras, 407, 2131,
  \dodoi{10.1111/j.1365-2966.2010.17028.x}

\bibitem[{{Dunkley} {et~al.}(2009){Dunkley}, {Komatsu}, {Nolta}, {Spergel},
  {Larson}, {Hinshaw}, {Page}, {Bennett}, {Gold}, {Jarosik}, {Weiland},
  {Halpern}, {Hill}, {Kogut}, {Limon}, {Meyer}, {Tucker}, {Wollack}, \&
  {Wright}}]{Dunkley2009}
{Dunkley}, J., {Komatsu}, E., {Nolta}, M.~R., {et~al.} 2009, \apjs, 180, 306,
  \dodoi{10.1088/0067-0049/180/2/306}

\bibitem[{{Farrow} {et~al.}(2015){Farrow}, {Cole}, {Norberg}, {Metcalfe},
  {Baldry}, {Bland-Hawthorn}, {Brown}, {Hopkins}, {Lacey}, {Liske}, {Loveday},
  {Palamara}, {Robotham}, \& {Sridhar}}]{Farrow2015}
{Farrow}, D.~J., {Cole}, S., {Norberg}, P., {et~al.} 2015, \mnras, 454, 2120,
  \dodoi{10.1093/mnras/stv2075}

\bibitem[{{Foreman-Mackey} {et~al.}(2013){Foreman-Mackey}, {Hogg}, {Lang}, \&
  {Goodman}}]{emcee2013}
{Foreman-Mackey}, D., {Hogg}, D.~W., {Lang}, D., \& {Goodman}, J. 2013, \pasp,
  125, 306, \dodoi{10.1086/670067}

\bibitem[{{Garilli} {et~al.}(2014){Garilli}, {Guzzo}, {Scodeggio},
  {Bolzonella}, {Abbas}, {Adami}, {Arnouts}, {Bel}, {Bottini}, {Branchini},
  {Cappi}, {Coupon}, {Cucciati}, {Davidzon}, {De Lucia}, {de la Torre},
  {Franzetti}, {Fritz}, {Fumana}, {Granett}, {Ilbert}, {Iovino}, {Krywult}, {Le
  Brun}, {Le F{\`e}vre}, {Maccagni}, {Ma{\l}ek}, {Marulli}, {McCracken},
  {Paioro}, {Polletta}, {Pollo}, {Schlagenhaufer}, {Tasca}, {Tojeiro},
  {Vergani}, {Zamorani}, {Zanichelli}, {Burden}, {Di Porto}, {Marchetti},
  {Marinoni}, {Mellier}, {Moscardini}, {Nichol}, {Peacock}, {Percival},
  {Phleps}, \& {Wolk}}]{Garilli2014}
{Garilli}, B., {Guzzo}, L., {Scodeggio}, M., {et~al.} 2014, \aap, 562, A23,
  \dodoi{10.1051/0004-6361/201322790}

\bibitem[{{Greene} {et~al.}(2022){Greene}, {Bezanson}, {Ouchi}, {Silverman}, \&
  {the PFS Galaxy Evolution Working Group}}]{PFS-galaxysurvey}
{Greene}, J., {Bezanson}, R., {Ouchi}, M., {Silverman}, J., \& {the PFS Galaxy
  Evolution Working Group}. 2022, arXiv e-prints, arXiv:2206.14908,
  \dodoi{10.48550/arXiv.2206.14908}

\bibitem[{{Hamilton}(1993)}]{Hamilton1993}
{Hamilton}, A.~J.~S. 1993, \apj, 417, 19, \dodoi{10.1086/173288}

\bibitem[{{Harris} {et~al.}(2020){Harris}, {Millman}, {van der Walt},
  {Gommers}, {Virtanen}, {Cournapeau}, {Wieser}, {Taylor}, {Berg}, {Smith},
  {Kern}, {Picus}, {Hoyer}, {van Kerkwijk}, {Brett}, {Haldane}, {del R{\'\i}o},
  {Wiebe}, {Peterson}, {G{\'e}rard-Marchant}, {Sheppard}, {Reddy}, {Weckesser},
  {Abbasi}, {Gohlke}, \& {Oliphant}}]{numpy2020}
{Harris}, C.~R., {Millman}, K.~J., {van der Walt}, S.~J., {et~al.} 2020, \nat,
  585, 357, \dodoi{10.1038/s41586-020-2649-2}

\bibitem[{{Hawkins} {et~al.}(2003){Hawkins}, {Maddox}, {Cole}, {Lahav},
  {Madgwick}, {Norberg}, {Peacock}, {Baldry}, {Baugh}, {Bland-Hawthorn},
  {Bridges}, {Cannon}, {Colless}, {Collins}, {Couch}, {Dalton}, {De Propris},
  {Driver}, {Efstathiou}, {Ellis}, {Frenk}, {Glazebrook}, {Jackson}, {Jones},
  {Lewis}, {Lumsden}, {Percival}, {Peterson}, {Sutherland}, \&
  {Taylor}}]{Hawkins2003}
{Hawkins}, E., {Maddox}, S., {Cole}, S., {et~al.} 2003, \mnras, 346, 78,
  \dodoi{10.1046/j.1365-2966.2003.07063.x}

\bibitem[{{Hearin} \& {Watson}(2013)}]{Hearin2013}
{Hearin}, A.~P., \& {Watson}, D.~F. 2013, \mnras, 435, 1313,
  \dodoi{10.1093/mnras/stt1374}

\bibitem[{{Hearin} {et~al.}(2014){Hearin}, {Watson}, {Becker}, {Reyes},
  {Berlind}, \& {Zentner}}]{Hearin2014}
{Hearin}, A.~P., {Watson}, D.~F., {Becker}, M.~R., {et~al.} 2014, \mnras, 444,
  729, \dodoi{10.1093/mnras/stu1443}

\bibitem[{Hunter(2007)}]{Matplotlib2007}
Hunter, J.~D. 2007, Computing in Science \& Engineering, 9, 90,
  \dodoi{10.1109/MCSE.2007.55}

\bibitem[{{Ilbert} {et~al.}(2005){Ilbert}, {Tresse}, {Zucca}, {Bardelli},
  {Arnouts}, {Zamorani}, {Pozzetti}, {Bottini}, {Garilli}, {Le Brun}, {Le
  F{\`e}vre}, {Maccagni}, {Picat}, {Scaramella}, {Scodeggio}, {Vettolani},
  {Zanichelli}, {Adami}, {Arnaboldi}, {Bolzonella}, {Cappi}, {Charlot},
  {Contini}, {Foucaud}, {Franzetti}, {Gavignaud}, {Guzzo}, {Iovino},
  {McCracken}, {Marano}, {Marinoni}, {Mathez}, {Mazure}, {Meneux}, {Merighi},
  {Paltani}, {Pello}, {Pollo}, {Radovich}, {Bondi}, {Bongiorno}, {Busarello},
  {Ciliegi}, {Lamareille}, {Mellier}, {Merluzzi}, {Ripepi}, \&
  {Rizzo}}]{Ilbert2005}
{Ilbert}, O., {Tresse}, L., {Zucca}, E., {et~al.} 2005, \aap, 439, 863,
  \dodoi{10.1051/0004-6361:20041961}

\bibitem[{{Ilbert} {et~al.}(2006){Ilbert}, {Arnouts}, {McCracken},
  {Bolzonella}, {Bertin}, {Le F{\`e}vre}, {Mellier}, {Zamorani}, {Pell{\`o}},
  {Iovino}, {Tresse}, {Le Brun}, {Bottini}, {Garilli}, {Maccagni}, {Picat},
  {Scaramella}, {Scodeggio}, {Vettolani}, {Zanichelli}, {Adami}, {Bardelli},
  {Cappi}, {Charlot}, {Ciliegi}, {Contini}, {Cucciati}, {Foucaud}, {Franzetti},
  {Gavignaud}, {Guzzo}, {Marano}, {Marinoni}, {Mazure}, {Meneux}, {Merighi},
  {Paltani}, {Pollo}, {Pozzetti}, {Radovich}, {Zucca}, {Bondi}, {Bongiorno},
  {Busarello}, {de La Torre}, {Gregorini}, {Lamareille}, {Mathez}, {Merluzzi},
  {Ripepi}, {Rizzo}, \& {Vergani}}]{Ilbert2006}
{Ilbert}, O., {Arnouts}, S., {McCracken}, H.~J., {et~al.} 2006, \aap, 457, 841,
  \dodoi{10.1051/0004-6361:20065138}

\bibitem[{{Jing} {et~al.}(1998){Jing}, {Mo}, \&
  {B{\"o}rner}}]{JingMoBoerner1998}
{Jing}, Y.~P., {Mo}, H.~J., \& {B{\"o}rner}, G. 1998, \apj, 494, 1,
  \dodoi{10.1086/305209}

\bibitem[{{Knobel} {et~al.}(2012){Knobel}, {Lilly}, {Iovino}, {Kova{\v{c}}},
  {Bschorr}, {Presotto}, {Oesch}, {Kampczyk}, {Carollo}, {Contini}, {Kneib},
  {Le Fevre}, {Mainieri}, {Renzini}, {Scodeggio}, {Zamorani}, {Bardelli},
  {Bolzonella}, {Bongiorno}, {Caputi}, {Cucciati}, {de la Torre}, {de Ravel},
  {Franzetti}, {Garilli}, {Lamareille}, {Le Borgne}, {Le Brun}, {Maier},
  {Mignoli}, {Pello}, {Peng}, {Perez Montero}, {Silverman}, {Tanaka}, {Tasca},
  {Tresse}, {Vergani}, {Zucca}, {Barnes}, {Bordoloi}, {Cappi}, {Cimatti},
  {Coppa}, {Koekemoer}, {L{\'o}pez-Sanjuan}, {McCracken}, {Moresco}, {Nair},
  {Pozzetti}, \& {Welikala}}]{Knobel2012}
{Knobel}, C., {Lilly}, S.~J., {Iovino}, A., {et~al.} 2012, \apj, 753, 121,
  \dodoi{10.1088/0004-637X/753/2/121}

\bibitem[{{Komatsu} {et~al.}(2009){Komatsu}, {Dunkley}, {Nolta}, {Bennett},
  {Gold}, {Hinshaw}, {Jarosik}, {Larson}, {Limon}, {Page}, {Spergel},
  {Halpern}, {Hill}, {Kogut}, {Meyer}, {Tucker}, {Weiland}, {Wollack}, \&
  {Wright}}]{Komatsu2009}
{Komatsu}, E., {Dunkley}, J., {Nolta}, M.~R., {et~al.} 2009, \apjs, 180, 330,
  \dodoi{10.1088/0067-0049/180/2/330}

\bibitem[{{Laigle} {et~al.}(2016){Laigle}, {McCracken}, {Ilbert}, {Hsieh},
  {Davidzon}, {Capak}, {Hasinger}, {Silverman}, {Pichon}, {Coupon}, {Aussel},
  {Le Borgne}, {Caputi}, {Cassata}, {Chang}, {Civano}, {Dunlop}, {Fynbo},
  {Kartaltepe}, {Koekemoer}, {Le F{\`e}vre}, {Le Floc'h}, {Leauthaud}, {Lilly},
  {Lin}, {Marchesi}, {Milvang-Jensen}, {Salvato}, {Sanders}, {Scoville},
  {Smolcic}, {Stockmann}, {Taniguchi}, {Tasca}, {Toft}, {Vaccari}, \&
  {Zabl}}]{Laigle2016}
{Laigle}, C., {McCracken}, H.~J., {Ilbert}, O., {et~al.} 2016, \apjs, 224, 24,
  \dodoi{10.3847/0067-0049/224/2/24}

\bibitem[{{Lan} {et~al.}(2016){Lan}, {M{\'e}nard}, \& {Mo}}]{LanMenardMo2016}
{Lan}, T.-W., {M{\'e}nard}, B., \& {Mo}, H. 2016, \mnras, 459, 3998,
  \dodoi{10.1093/mnras/stw898}

\bibitem[{{Landy} \& {Szalay}(1993)}]{Landy1993}
{Landy}, S.~D., \& {Szalay}, A.~S. 1993, \apj, 412, 64, \dodoi{10.1086/172900}

\bibitem[{{Li} {et~al.}(2007){Li}, {Jing}, {Kauffmann}, {B{\"o}rner}, {Kang},
  \& {Wang}}]{Li-07}
{Li}, C., {Jing}, Y.~P., {Kauffmann}, G., {et~al.} 2007, \mnras, 376, 984,
  \dodoi{10.1111/j.1365-2966.2007.11518.x}

\bibitem[{{Li} {et~al.}(2006{\natexlab{a}}){Li}, {Jing}, {Kauffmann},
  {B{\"o}rner}, {White}, \& {Cheng}}]{Li-06c}
---. 2006{\natexlab{a}}, \mnras, 368, 37,
  \dodoi{10.1111/j.1365-2966.2006.10177.x}

\bibitem[{{Li} {et~al.}(2006{\natexlab{b}}){Li}, {Kauffmann}, {Jing}, {White},
  {B{\"o}rner}, \& {Cheng}}]{Li-06b}
{Li}, C., {Kauffmann}, G., {Jing}, Y.~P., {et~al.} 2006{\natexlab{b}}, \mnras,
  368, 21, \dodoi{10.1111/j.1365-2966.2006.10066.x}

\bibitem[{{Li} {et~al.}(2006{\natexlab{c}}){Li}, {Kauffmann}, {Wang}, {White},
  {Heckman}, \& {Jing}}]{Li-06a}
{Li}, C., {Kauffmann}, G., {Wang}, L., {et~al.} 2006{\natexlab{c}}, \mnras,
  373, 457, \dodoi{10.1111/j.1365-2966.2006.11079.x}

\bibitem[{{Li} \& {White}(2009)}]{Li2009}
{Li}, C., \& {White}, S. D.~M. 2009, \mnras, 398, 2177,
  \dodoi{10.1111/j.1365-2966.2009.15268.x}

\bibitem[{{Lilly} {et~al.}(2007){Lilly}, {Le F{\`e}vre}, {Renzini}, {Zamorani},
  {Scodeggio}, {Contini}, {Carollo}, {Hasinger}, {Kneib}, {Iovino}, {Le Brun},
  {Maier}, {Mainieri}, {Mignoli}, {Silverman}, {Tasca}, {Bolzonella},
  {Bongiorno}, {Bottini}, {Capak}, {Caputi}, {Cimatti}, {Cucciati}, {Daddi},
  {Feldmann}, {Franzetti}, {Garilli}, {Guzzo}, {Ilbert}, {Kampczyk}, {Kovac},
  {Lamareille}, {Leauthaud}, {Le Borgne}, {McCracken}, {Marinoni}, {Pello},
  {Ricciardelli}, {Scarlata}, {Vergani}, {Sanders}, {Schinnerer}, {Scoville},
  {Taniguchi}, {Arnouts}, {Aussel}, {Bardelli}, {Brusa}, {Cappi}, {Ciliegi},
  {Finoguenov}, {Foucaud}, {Franceschini}, {Halliday}, {Impey}, {Knobel},
  {Koekemoer}, {Kurk}, {Maccagni}, {Maddox}, {Marano}, {Marconi}, {Meneux},
  {Mobasher}, {Moreau}, {Peacock}, {Porciani}, {Pozzetti}, {Scaramella},
  {Schiminovich}, {Shopbell}, {Smail}, {Thompson}, {Tresse}, {Vettolani},
  {Zanichelli}, \& {Zucca}}]{Lilly2007}
{Lilly}, S.~J., {Le F{\`e}vre}, O., {Renzini}, A., {et~al.} 2007, \apjs, 172,
  70, \dodoi{10.1086/516589}

\bibitem[{{Lu} {et~al.}(2011){Lu}, {Mo}, {Weinberg}, \& {Katz}}]{LuYu_etal2011}
{Lu}, Y., {Mo}, H.~J., {Weinberg}, M.~D., \& {Katz}, N. 2011, \mnras, 416,
  1949, \dodoi{10.1111/j.1365-2966.2011.19170.x}

\bibitem[{{Lu} {et~al.}(2014){Lu}, {Mo}, {Lu}, {Katz}, {Weinberg}, {van den
  Bosch}, \& {Yang}}]{Lu2014a}
{Lu}, Z., {Mo}, H.~J., {Lu}, Y., {et~al.} 2014, \mnras, 439, 1294,
  \dodoi{10.1093/mnras/stu016}

\bibitem[{{Lu} {et~al.}(2015){Lu}, {Mo}, {Lu}, {Katz}, {Weinberg}, {van den
  Bosch}, \& {Yang}}]{LuZ_etal2015}
---. 2015, \mnras, 450, 1604, \dodoi{10.1093/mnras/stv667}

\bibitem[{{Madgwick} {et~al.}(2003){Madgwick}, {Hawkins}, {Lahav}, {Maddox},
  {Norberg}, {Peacock}, {Baldry}, {Baugh}, {Bland-Hawthorn}, {Bridges},
  {Cannon}, {Cole}, {Colless}, {Collins}, {Couch}, {Dalton}, {De Propris},
  {Driver}, {Efstathiou}, {Ellis}, {Frenk}, {Glazebrook}, {Jackson}, {Lewis},
  {Lumsden}, {Peterson}, {Sutherland}, \& {Taylor}}]{Madgwick2003}
{Madgwick}, D.~S., {Hawkins}, E., {Lahav}, O., {et~al.} 2003, \mnras, 344, 847,
  \dodoi{10.1046/j.1365-8711.2003.06861.x}

\bibitem[{{Marulli} {et~al.}(2013){Marulli}, {Bolzonella}, {Branchini},
  {Davidzon}, {de la Torre}, {Granett}, {Guzzo}, {Iovino}, {Moscardini},
  {Pollo}, {Abbas}, {Adami}, {Arnouts}, {Bel}, {Bottini}, {Cappi}, {Coupon},
  {Cucciati}, {De Lucia}, {Fritz}, {Franzetti}, {Fumana}, {Garilli}, {Ilbert},
  {Krywult}, {Le Brun}, {Le F{\`e}vre}, {Maccagni}, {Ma{\l}ek}, {McCracken},
  {Paioro}, {Polletta}, {Schlagenhaufer}, {Scodeggio}, {Tasca}, {Tojeiro},
  {Vergani}, {Zanichelli}, {Burden}, {Di Porto}, {Marchetti}, {Marinoni},
  {Mellier}, {Nichol}, {Peacock}, {Percival}, {Phleps}, {Wolk}, \&
  {Zamorani}}]{Marulli2013}
{Marulli}, F., {Bolzonella}, M., {Branchini}, E., {et~al.} 2013, \aap, 557,
  A17, \dodoi{10.1051/0004-6361/201321476}

\bibitem[{{Meneux} {et~al.}(2008){Meneux}, {Guzzo}, {Garilli}, {Le F{\`e}vre},
  {Pollo}, {Blaizot}, {De Lucia}, {Bolzonella}, {Lamareille}, {Pozzetti},
  {Cappi}, {Iovino}, {Marinoni}, {McCracken}, {de la Torre}, {Bottini}, {Le
  Brun}, {Maccagni}, {Picat}, {Scaramella}, {Scodeggio}, {Tresse}, {Vettolani},
  {Zanichelli}, {Abbas}, {Adami}, {Arnouts}, {Bardelli}, {Bongiorno},
  {Charlot}, {Ciliegi}, {Contini}, {Cucciati}, {Foucaud}, {Franzetti},
  {Gavignaud}, {Ilbert}, {Marano}, {Mazure}, {Merighi}, {Paltani}, {Pell{\`o}},
  {Radovich}, {Vergani}, {Zamorani}, \& {Zucca}}]{Meneux2008}
{Meneux}, B., {Guzzo}, L., {Garilli}, B., {et~al.} 2008, \aap, 478, 299,
  \dodoi{10.1051/0004-6361:20078182}

\bibitem[{{Meneux} {et~al.}(2009){Meneux}, {Guzzo}, {de la Torre}, {Porciani},
  {Zamorani}, {Abbas}, {Bolzonella}, {Garilli}, {Iovino}, {Pozzetti}, {Zucca},
  {Lilly}, {Le F{\`e}vre}, {Kneib}, {Carollo}, {Contini}, {Mainieri},
  {Renzini}, {Scodeggio}, {Bardelli}, {Bongiorno}, {Caputi}, {Coppa},
  {Cucciati}, {de Ravel}, {Franzetti}, {Kampczyk}, {Knobel}, {Kova{\v{c}}},
  {Lamareille}, {Le Borgne}, {Le Brun}, {Maier}, {Pell{\`o}}, {Peng}, {Perez
  Montero}, {Ricciardelli}, {Silverman}, {Tanaka}, {Tasca}, {Tresse},
  {Vergani}, {Bottini}, {Cappi}, {Cimatti}, {Cassata}, {Fumana}, {Koekemoer},
  {Leauthaud}, {Maccagni}, {Marinoni}, {McCracken}, {Memeo}, {Oesch}, \&
  {Scaramella}}]{Meneux2009}
{Meneux}, B., {Guzzo}, L., {de la Torre}, S., {et~al.} 2009, \aap, 505, 463,
  \dodoi{10.1051/0004-6361/200912314}

\bibitem[{Meng {et~al.}(2023)Meng, Li, Mo, Chen, \& Wang}]{Mock_data}
Meng, J., Li, C., Mo, H., Chen, Y., \& Wang, K. 2023, {The mock samples for
  high-z spectroscopic galaxy survey},  Zenodo, \dodoi{10.5281/zenodo.10113272}

\bibitem[{{Mo} {et~al.}(2010){Mo}, {van den Bosch}, \&
  {White}}]{MoBoschWhite2010}
{Mo}, H., {van den Bosch}, F.~C., \& {White}, S. 2010, {Galaxy Formation and
  Evolution}

\bibitem[{{Mohammad} {et~al.}(2018){Mohammad}, {Bianchi}, {Percival}, {de la
  Torre}, {Guzzo}, {Granett}, {Branchini}, {Bolzonella}, {Garilli},
  {Scodeggio}, {Abbas}, {Adami}, {Bel}, {Bottini}, {Cappi}, {Cucciati},
  {Davidzon}, {Franzetti}, {Fritz}, {Iovino}, {Krywult}, {Le Brun}, {Le
  F{\`e}vre}, {Ma{\l}ek}, {Marulli}, {Polletta}, {Pollo}, {Tasca}, {Tojeiro},
  {Vergani}, {Zanichelli}, {Arnouts}, {Coupon}, {De Lucia}, {Ilbert},
  {Moscardini}, \& {Moutard}}]{Mohammad2018}
{Mohammad}, F.~G., {Bianchi}, D., {Percival}, W.~J., {et~al.} 2018, \aap, 619,
  A17, \dodoi{10.1051/0004-6361/201833853}

\bibitem[{{Moster} {et~al.}(2011){Moster}, {Somerville}, {Newman}, \&
  {Rix}}]{Moster2011}
{Moster}, B.~P., {Somerville}, R.~S., {Newman}, J.~A., \& {Rix}, H.-W. 2011,
  \apj, 731, 113, \dodoi{10.1088/0004-637X/731/2/113}

\bibitem[{{Muzzin} {et~al.}(2013){Muzzin}, {Marchesini}, {Stefanon}, {Franx},
  {Milvang-Jensen}, {Dunlop}, {Fynbo}, {Brammer}, {Labb{\'e}}, \& {van
  Dokkum}}]{Muzzin2013}
{Muzzin}, A., {Marchesini}, D., {Stefanon}, M., {et~al.} 2013, \apjs, 206, 8,
  \dodoi{10.1088/0067-0049/206/1/8}

\bibitem[{{Norberg} {et~al.}(2001){Norberg}, {Baugh}, {Hawkins}, {Maddox},
  {Peacock}, {Cole}, {Frenk}, {Bland-Hawthorn}, {Bridges}, {Cannon}, {Colless},
  {Collins}, {Couch}, {Dalton}, {De Propris}, {Driver}, {Efstathiou}, {Ellis},
  {Glazebrook}, {Jackson}, {Lahav}, {Lewis}, {Lumsden}, {Madgwick}, {Peterson},
  {Sutherland}, \& {Taylor}}]{Norberg2001}
{Norberg}, P., {Baugh}, C.~M., {Hawkins}, E., {et~al.} 2001, \mnras, 328, 64,
  \dodoi{10.1046/j.1365-8711.2001.04839.x}

\bibitem[{{Parkinson} {et~al.}(2008){Parkinson}, {Cole}, \&
  {Helly}}]{Parkinson2008}
{Parkinson}, H., {Cole}, S., \& {Helly}, J. 2008, \mnras, 383, 557,
  \dodoi{10.1111/j.1365-2966.2007.12517.x}

\bibitem[{{Pearl} {et~al.}(2022){Pearl}, {Bezanson}, {Zentner}, {Newman},
  {Goulding}, {Whitaker}, {Johnson}, \& {Greene}}]{Pearl2022}
{Pearl}, A.~N., {Bezanson}, R., {Zentner}, A.~R., {et~al.} 2022, \apj, 925,
  180, \dodoi{10.3847/1538-4357/ac3fb5}

\bibitem[{{Pollo} {et~al.}(2006){Pollo}, {Guzzo}, {Le F{\`e}vre}, {Meneux},
  {Cappi}, {Franzetti}, {Iovino}, {McCracken}, {Marinoni}, {Zamorani},
  {Bottini}, {Garilli}, {Le Brun}, {Maccagni}, {Picat}, {Scaramella},
  {Scodeggio}, {Tresse}, {Vettolani}, {Zanichelli}, {Adami}, {Arnouts},
  {Bardelli}, {Bolzonella}, {Charlot}, {Ciliegi}, {Contini}, {Foucaud},
  {Gavignaud}, {Ilbert}, {Marano}, {Mazure}, {Merighi}, {Paltani}, {Pell{\`o}},
  {Pozzetti}, {Radovich}, {Zucca}, {Bondi}, {Bongiorno}, {Busarello},
  {Cucciati}, {Gregorini}, {Lamareille}, {Mathez}, {Mellier}, {Merluzzi},
  {Ripepi}, \& {Rizzo}}]{Pollo2006}
{Pollo}, A., {Guzzo}, L., {Le F{\`e}vre}, O., {et~al.} 2006, \aap, 451, 409,
  \dodoi{10.1051/0004-6361:20054705}

\bibitem[{{Pozzetti} {et~al.}(2007){Pozzetti}, {Bolzonella}, {Lamareille},
  {Zamorani}, {Franzetti}, {Le F{\`e}vre}, {Iovino}, {Temporin}, {Ilbert},
  {Arnouts}, {Charlot}, {Brinchmann}, {Zucca}, {Tresse}, {Scodeggio}, {Guzzo},
  {Bottini}, {Garilli}, {Le Brun}, {Maccagni}, {Picat}, {Scaramella},
  {Vettolani}, {Zanichelli}, {Adami}, {Bardelli}, {Cappi}, {Ciliegi},
  {Contini}, {Foucaud}, {Gavignaud}, {McCracken}, {Marano}, {Marinoni},
  {Mazure}, {Meneux}, {Merighi}, {Paltani}, {Pell{\`o}}, {Pollo}, {Radovich},
  {Bondi}, {Bongiorno}, {Cucciati}, {de la Torre}, {Gregorini}, {Mellier},
  {Merluzzi}, {Vergani}, \& {Walcher}}]{Pozzetti2007}
{Pozzetti}, L., {Bolzonella}, M., {Lamareille}, F., {et~al.} 2007, \aap, 474,
  443, \dodoi{10.1051/0004-6361:20077609}

\bibitem[{{Pozzetti} {et~al.}(2010){Pozzetti}, {Bolzonella}, {Zucca},
  {Zamorani}, {Lilly}, {Renzini}, {Moresco}, {Mignoli}, {Cassata}, {Tasca},
  {Lamareille}, {Maier}, {Meneux}, {Halliday}, {Oesch}, {Vergani}, {Caputi},
  {Kova{\v c}}, {Cimatti}, {Cucciati}, {Iovino}, {Peng}, {Carollo}, {Contini},
  {Kneib}, {Le F{\'e}vre}, {Mainieri}, {Scodeggio}, {Bardelli}, {Bongiorno},
  {Coppa}, {de la Torre}, {de Ravel}, {Franzetti}, {Garilli}, {Kampczyk},
  {Knobel}, {Le Borgne}, {Le Brun}, {Pell{\`o}}, {Perez Montero},
  {Ricciardelli}, {Silverman}, {Tanaka}, {Tresse}, {Abbas}, {Bottini}, {Cappi},
  {Guzzo}, {Koekemoer}, {Leauthaud}, {Maccagni}, {Marinoni}, {McCracken},
  {Memeo}, {Porciani}, {Scaramella}, {Scarlata}, \& {Scoville}}]{Pozzetti2010}
{Pozzetti}, L., {Bolzonella}, M., {Zucca}, E., {et~al.} 2010, A\&A, 523, A13,
  \dodoi{10.1051/0004-6361/200913020}

\bibitem[{{Schmidt}(1968)}]{Schmidt1968}
{Schmidt}, M. 1968, ApJ, 151, 393, \dodoi{10.1086/149446}

\bibitem[{{Scodeggio} {et~al.}(2018){Scodeggio}, {Guzzo}, {Garilli}, {Granett},
  {Bolzonella}, {de la Torre}, {Abbas}, {Adami}, {Arnouts}, {Bottini}, {Cappi},
  {Coupon}, {Cucciati}, {Davidzon}, {Franzetti}, {Fritz}, {Iovino}, {Krywult},
  {Le Brun}, {Le F{\`e}vre}, {Maccagni}, {Ma{\l}ek}, {Marchetti}, {Marulli},
  {Polletta}, {Pollo}, {Tasca}, {Tojeiro}, {Vergani}, {Zanichelli}, {Bel},
  {Branchini}, {De Lucia}, {Ilbert}, {McCracken}, {Moutard}, {Peacock},
  {Zamorani}, {Burden}, {Fumana}, {Jullo}, {Marinoni}, {Mellier}, {Moscardini},
  \& {Percival}}]{Scodeggio2018}
{Scodeggio}, M., {Guzzo}, L., {Garilli}, B., {et~al.} 2018, \aap, 609, A84,
  \dodoi{10.1051/0004-6361/201630114}

\bibitem[{{Shimono} {et~al.}(2016){Shimono}, {Tamura}, {Takato}, {Yasuda},
  {Suzuki}, {Loomis}, {Lupton}, {Moritani}, \& {Yabe}}]{Shimono2016}
{Shimono}, A., {Tamura}, N., {Takato}, N., {et~al.} 2016, in Society of
  Photo-Optical Instrumentation Engineers (SPIE) Conference Series, Vol. 9913,
  Software and Cyberinfrastructure for Astronomy IV, 99133B,
  \dodoi{10.1117/12.2232844}

\bibitem[{{Somerville} {et~al.}(2004){Somerville}, {Lee}, {Ferguson},
  {Gardner}, {Moustakas}, \& {Giavalisco}}]{Somerville2004}
{Somerville}, R.~S., {Lee}, K., {Ferguson}, H.~C., {et~al.} 2004, \apjl, 600,
  L171, \dodoi{10.1086/378628}

\bibitem[{{Springel} {et~al.}(2005){Springel}, {White}, {Jenkins}, {Frenk},
  {Yoshida}, {Gao}, {Navarro}, {Thacker}, {Croton}, {Helly}, {Peacock}, {Cole},
  {Thomas}, {Couchman}, {Evrard}, {Colberg}, \& {Pearce}}]{Springel2005}
{Springel}, V., {White}, S. D.~M., {Jenkins}, A., {et~al.} 2005, \nat, 435,
  629, \dodoi{10.1038/nature03597}

\bibitem[{{Springel} {et~al.}(2018){Springel}, {Pakmor}, {Pillepich},
  {Weinberger}, {Nelson}, {Hernquist}, {Vogelsberger}, {Genel}, {Torrey},
  {Marinacci}, \& {Naiman}}]{Springel_etal2018}
{Springel}, V., {Pakmor}, R., {Pillepich}, A., {et~al.} 2018, \mnras, 475, 676,
  \dodoi{10.1093/mnras/stx3304}

\bibitem[{{Sunayama} {et~al.}(2019){Sunayama}, {Takada}, {Reinecke}, {Makiya},
  {Nishimichi}, {Komatsu}, {Saito}, {Tamura}, \& {Yabe}}]{Sunayama2019}
{Sunayama}, T., {Takada}, M., {Reinecke}, M., {et~al.} 2019, arXiv e-prints,
  arXiv:1912.06583.
\newblock \doarXiv{1912.06583}

\bibitem[{{Takada} {et~al.}(2014){Takada}, {Ellis}, {Chiba}, {Greene},
  {Aihara}, {Arimoto}, {Bundy}, {Cohen}, {Dor{\'e}}, {Graves}, {Gunn},
  {Heckman}, {Hirata}, {Ho}, {Kneib}, {Le F{\`e}vre}, {Lin}, {More},
  {Murayama}, {Nagao}, {Ouchi}, {Seiffert}, {Silverman}, {Sodr{\'e}},
  {Spergel}, {Strauss}, {Sugai}, {Suto}, {Takami}, \& {Wyse}}]{Takada2014}
{Takada}, M., {Ellis}, R.~S., {Chiba}, M., {et~al.} 2014, \pasj, 66, R1,
  \dodoi{10.1093/pasj/pst019}

\bibitem[{{Virtanen} {et~al.}(2020){Virtanen}, {Gommers}, {Oliphant},
  {Haberland}, {Reddy}, {Cournapeau}, {Burovski}, {Peterson}, {Weckesser},
  {Bright}, {van der Walt}, {Brett}, {Wilson}, {Millman}, {Mayorov}, {Nelson},
  {Jones}, {Kern}, {Larson}, {Carey}, {Polat}, {Feng}, {Moore}, {VanderPlas},
  {Laxalde}, {Perktold}, {Cimrman}, {Henriksen}, {Quintero}, {Harris},
  {Archibald}, {Ribeiro}, {Pedregosa}, {van Mulbregt}, \& {SciPy 1. 0
  Contributors}}]{scipy2020}
{Virtanen}, P., {Gommers}, R., {Oliphant}, T.~E., {et~al.} 2020, Nature
  Methods, 17, 261, \dodoi{10.1038/s41592-019-0686-2}

\bibitem[{{Wang} {et~al.}(2016){Wang}, {Mo}, {Yang}, {Zhang}, {Shi}, {Jing},
  {Liu}, {Li}, {Kang}, \& {Gao}}]{Wang2016}
{Wang}, H., {Mo}, H.~J., {Yang}, X., {et~al.} 2016, ApJ, 831, 164,
  \dodoi{10.3847/0004-637X/831/2/164}

\bibitem[{{Wang} {et~al.}(2023){Wang}, {Mo}, {Li}, \& {Chen}}]{Wang2023}
{Wang}, K., {Mo}, H., {Li}, C., \& {Chen}, Y. 2023, \mnras, 520, 1774,
  \dodoi{10.1093/mnras/stad262}

\bibitem[{{Wang} {et~al.}(2020){Wang}, {Mo}, {Li}, {Meng}, \& {Chen}}]{Wang-20}
{Wang}, K., {Mo}, H.~J., {Li}, C., {Meng}, J., \& {Chen}, Y. 2020, \mnras, 499,
  89, \dodoi{10.1093/mnras/staa2816}

\bibitem[{{Weaver} {et~al.}(2022{\natexlab{a}}){Weaver}, {Davidzon}, {Toft},
  {Ilbert}, {McCracken}, {Gould}, {Jespersen}, {Steinhardt}, {Lagos}, {Capak},
  {Casey}, {Chartab}, {Faisst}, {Hayward}, {Kartaltepe}, {Kauffmann},
  {Koekemoer}, {Kokorev}, {Laigle}, {Liu}, {Long}, {Magdis}, {McPartland},
  {Milvang-Jensen}, {Mobasher}, {Moneti}, {Peng}, {Sanders}, {Shuntov},
  {Sneppen}, {Valentino}, {Zalesky}, \& {Zamorani}}]{Weaver2022gsmf}
{Weaver}, J.~R., {Davidzon}, I., {Toft}, S., {et~al.} 2022{\natexlab{a}}, arXiv
  e-prints, arXiv:2212.02512, \dodoi{10.48550/arXiv.2212.02512}

\bibitem[{{Weaver} {et~al.}(2022{\natexlab{b}}){Weaver}, {Kauffmann}, {Ilbert},
  {McCracken}, {Moneti}, {Toft}, {Brammer}, {Shuntov}, {Davidzon}, {Hsieh},
  {Laigle}, {Anastasiou}, {Jespersen}, {Vinther}, {Capak}, {Casey},
  {McPartland}, {Milvang-Jensen}, {Mobasher}, {Sanders}, {Zalesky}, {Arnouts},
  {Aussel}, {Dunlop}, {Faisst}, {Franx}, {Furtak}, {Fynbo}, {Gould}, {Greve},
  {Gwyn}, {Kartaltepe}, {Kashino}, {Koekemoer}, {Kokorev}, {Le F{\`e}vre},
  {Lilly}, {Masters}, {Magdis}, {Mehta}, {Peng}, {Riechers}, {Salvato},
  {Sawicki}, {Scarlata}, {Scoville}, {Shirley}, {Silverman}, {Sneppen},
  {Smolc̆i{\'c}}, {Steinhardt}, {Stern}, {Tanaka}, {Taniguchi}, {Teplitz},
  {Vaccari}, {Wang}, \& {Zamorani}}]{Weaver2022}
{Weaver}, J.~R., {Kauffmann}, O.~B., {Ilbert}, O., {et~al.} 2022{\natexlab{b}},
  \apjs, 258, 11, \dodoi{10.3847/1538-4365/ac3078}

\bibitem[{{Wechsler} {et~al.}(2002){Wechsler}, {Bullock}, {Primack},
  {Kravtsov}, \& {Dekel}}]{Wechsler2002}
{Wechsler}, R.~H., {Bullock}, J.~S., {Primack}, J.~R., {Kravtsov}, A.~V., \&
  {Dekel}, A. 2002, \apj, 568, 52, \dodoi{10.1086/338765}

\bibitem[{{Wechsler} \& {Tinker}(2018)}]{WechslerTinker2018}
{Wechsler}, R.~H., \& {Tinker}, J.~L. 2018, \araa, 56, 435,
  \dodoi{10.1146/annurev-astro-081817-051756}

\bibitem[{{White} \& {Rees}(1978)}]{White1978}
{White}, S.~D.~M., \& {Rees}, M.~J. 1978, \mnras, 183, 341,
  \dodoi{10.1093/mnras/183.3.341}

\bibitem[{{Yang} {et~al.}(2003){Yang}, {Mo}, \& {van den
  Bosch}}]{YangMoBosch2003}
{Yang}, X., {Mo}, H.~J., \& {van den Bosch}, F.~C. 2003, \mnras, 339, 1057,
  \dodoi{10.1046/j.1365-8711.2003.06254.x}

\bibitem[{{York} {et~al.}(2000){York}, {Adelman}, {Anderson}, {Anderson},
  {Annis}, {Bahcall}, {Bakken}, {Barkhouser}, \& {et al.}}]{York-00}
{York}, D.~G., {Adelman}, J., {Anderson}, John~E., J., {et~al.} 2000, \aj, 120,
  1579, \dodoi{10.1086/301513}

\bibitem[{{Zehavi} {et~al.}(2005){Zehavi}, {Zheng}, {Weinberg}, {Frieman},
  {Berlind}, {Blanton}, {Scoccimarro}, {Sheth}, {Strauss}, {Kayo}, {Suto},
  {Fukugita}, {Nakamura}, {Bahcall}, {Brinkmann}, {Gunn}, {Hennessy},
  {Ivezi{\'c}}, {Knapp}, {Loveday}, {Meiksin}, {Schlegel}, {Schneider},
  {Szapudi}, {Tegmark}, {Vogeley}, {York}, \& {SDSS
  Collaboration}}]{Zehavi2005}
{Zehavi}, I., {Zheng}, Z., {Weinberg}, D.~H., {et~al.} 2005, \apj, 630, 1,
  \dodoi{10.1086/431891}

\bibitem[{{Zehavi} {et~al.}(2011){Zehavi}, {Zheng}, {Weinberg}, {Blanton},
  {Bahcall}, {Berlind}, {Brinkmann}, {Frieman}, {Gunn}, {Lupton}, {Nichol},
  {Percival}, {Schneider}, {Skibba}, {Strauss}, {Tegmark}, \&
  {York}}]{Zehavi_etal2011}
---. 2011, \apj, 736, 59, \dodoi{10.1088/0004-637X/736/1/59}

\bibitem[{{Zheng} {et~al.}(2005){Zheng}, {Berlind}, {Weinberg}, {Benson},
  {Baugh}, {Cole}, {Dav{\'e}}, {Frenk}, {Katz}, \& {Lacey}}]{Zheng2005}
{Zheng}, Z., {Berlind}, A.~A., {Weinberg}, D.~H., {et~al.} 2005, \apj, 633,
  791, \dodoi{10.1086/466510}

\bibitem[{{Zucca} {et~al.}(2009){Zucca}, {Bardelli}, {Bolzonella}, {Zamorani},
  {Ilbert}, {Pozzetti}, {Mignoli}, {Kova{\v{c}}}, {Lilly}, \&
  {Tresse}}]{Zucca2009}
{Zucca}, E., {Bardelli}, S., {Bolzonella}, M., {et~al.} 2009, \aap, 508, 1217,
  \dodoi{10.1051/0004-6361/200912665}

\end{thebibliography}
\bibliographystyle{aasjournal}



\end{document}